\documentclass[11pt]{article}
\pdfoutput = 1
\usepackage{jcapmod}

\usepackage{booktabs}
\usepackage[english]{babel}
\usepackage{amsmath,amssymb,amsbsy,amstext, amsthm, simplewick, amsfonts}
\usepackage{graphicx}
\usepackage[small]{caption}
\usepackage{siunitx}
\usepackage{upgreek}
\usepackage{framed}
\usepackage{wrapfig}
\usepackage{multirow}
\usepackage{subfig}
\usepackage{bbm}
\usepackage[svgnames,dvipsnames,x11names]{xcolor}
\usepackage[utf8x]{inputenc}
\usepackage{selinput}
\usepackage{bm}
\usepackage{float}

\usepackage{colortbl}
\definecolor{lightgreen}{cmyk}{0.2, 0, 0.2, 0.2}
\definecolor{lightgray}{cmyk}{0.1,0.2,0,0.1}
\definecolor{lightgray2}{cmyk}{0.1,0.1,0,0.1}

\makeatletter
\newlength{\apb@width}
\newcommand{\autoparbox}[2][c]{\settowidth{\apb@width}{#2}\parbox[#1]{\apb@width}{#2}}
\newcommand{\includegraphicsbox}[2][]{\autoparbox{\includegraphics[#1]{#2}}}
\makeatother

\def\d{{\rm d}}

\def\k{{\bf k}}
\def\s{{\bf s}}

\def\n{{\bf n}}
\def\r{{\bf r}}
\def\q{{\bf q}}

\def\x{{\bf x}}

\def\H{{\sf H}}
\def\G{{\sf G}}
\def\K{{\sf K}}
\def\I{{\cal I}}
\def\J{{\sf J}}

\def\fnl{f_{\rm NL}}

\def\hs{\hskip 1pt}
\def\bhs{\hskip -0.5pt}

\newcommand{\be}{\begin{equation}}
\newcommand{\ee}{\end{equation}}

\def\bk{{\bf k}}
\def\bq{{\bf q}}

\def\eps{\boldsymbol\varepsilon}

\begin{document}

\begin{titlepage}
\setcounter{page}{1} \baselineskip=15.5pt \thispagestyle{empty}
\bigskip\

\vspace{1cm}
\begin{center}
{\fontsize{20}{24}\selectfont  \sffamily \bfseries  Galaxy Bispectrum from}\\[12pt]
{\fontsize{20}{24}\selectfont  \sffamily \bfseries  Massive Spinning Particles}
\end{center}

\vspace{0.2cm}
\begin{center}
{\fontsize{13}{30}\selectfont  Azadeh Moradinezhad~Dizgah$^{1}$, Hayden Lee$^{1,\hskip 1pt2}$, Julian B.~Mu\~noz$^{1}$, and Cora Dvorkin$^{1}$}
\end{center}

\begin{center}
\vskip 8pt
\textsl{$^{1}$ 
Department of Physics, Harvard University,\\ 17 Oxford Street, Cambridge, MA 02138, USA}
\vskip 8pt
\textsl{$^{2}$ Institute for Advanced Study, \\ Hong Kong University of Science and Technology, Hong Kong}  
\end{center}

\vspace{1.2cm}
\hrule \vspace{0.3cm}
\noindent {\sffamily \bfseries Abstract} \\[0.1cm]
Massive spinning particles, if present during inflation, lead to a distinctive bispectrum of primordial perturbations, the shape and amplitude of which depend on the masses and spins of the extra particles. This signal, in turn, leaves an imprint in the statistical distribution of galaxies; in particular, as a non-vanishing galaxy bispectrum, which can be used to probe the masses and spins of these particles. In this paper, we present for the first time a new theoretical template for the bispectrum generated by massive spinning particles, valid for a general triangle configuration. We then proceed to perform a Fisher-matrix forecast to assess the potential of two next-generation spectroscopic galaxy surveys, EUCLID and DESI, to constrain the primordial non-Gaussianity sourced by these extra particles. We model the galaxy bispectrum using tree-level perturbation theory, accounting for redshift-space distortions and the Alcock-Paczynski effect, and forecast constraints on the primordial non-Gaussianity parameters marginalizing over all relevant biases and cosmological parameters. Our results suggest that these surveys would potentially be sensitive to any primordial non-Gaussianity with an amplitude larger than $f_{\rm NL}\approx 1$, for massive particles with spins 2, 3, and 4. Interestingly, if non-Gaussianities are present at that level, these surveys will be able to infer the masses of these spinning particles to within tens of percent. If detected, this would provide a very clear window into the particle content of our Universe during inflation. 

\vskip 10pt
\hrule
\vskip 10pt

\vspace{0.6cm}
\end{titlepage}
\tableofcontents
\newpage
\section{Introduction}
Understanding the origin of primordial fluctuations, which seed the large-scale structure of our Universe, is one of the fundamental open questions in cosmology. In the simplest models of inflation, characterized by a single degree of freedom~\cite{Guth:1980zm,Linde:1981mu,Albrecht:1982wi}, 
primordial fluctuations are produced by quantum fluctuations of a scalar field, the inflaton, as it slowly rolls down its potential \cite{Mukhanov:1981xt,Starobinsky:1982ee,Hawking:1982cz,Guth:1982ec}. Generically, these models predict adiabatic, superhorizon perturbations described by both a nearly scale-invariant spectrum and, to a good approximation, a Gaussian distribution \cite{Allen:1987vq,Falk:1992zg,Gangui:1993tt}. These predictions are in accordance with the latest measurements of the cosmic microwave background (CMB) anisotropies by the Planck satellite \cite{Ade:2015lrj,Ade:2015ava}.

However, deviations from the assumptions in the simplest inflationary models, such as non-slow-roll evolution,
non-Bunch-Davis vacuum states, non-canonical kinetic terms, or additional degrees of freedom during inflation, can generate significant departures from Gaussianity (see e.g.~Refs.~\cite{Bartolo:2004if, Chen:2010xka} for extensive reviews). 
In particular, the presence of extra particles during inflation---as generically expected from its ultraviolet completions~\cite{Baumann:2014nda}---can modify the distribution of primordial fluctuations~\cite{Chen:2009zp, Baumann:2011nk, Achucarro:2012sm, Noumi:2012vr, Arkani-Hamed:2015bza, Chen:2015lza, Dimastrogiovanni:2015pla, Flauger:2016idt, Lee:2016vti, Chen:2016hrz, Delacretaz:2016nhw, Kehagias:2017cym, Biagetti:2017viz, Kumar:2017ecc, Franciolini:2017ktv, Baumann:2017jvh,An:2017rwo}. 
This leaves distinct imprints on the primordial bispectrum, with an angular dependence determined by the spin of a new particle~\cite{Arkani-Hamed:2015bza, Lee:2016vti} and an oscillatory (or a power-law) feature determined by its mass~\cite{Chen:2009zp, Baumann:2011nk, Arkani-Hamed:2015bza}. 
Hence, the study of primordial non-Gaussianity (PNG) can explore the particle content of our Universe at the highest energies, inaccessible to accelerators.

The most stringent limits on primordial non-Gaussianity of several shapes are currently placed from measurements of the bispectrum of the CMB fluctuations~\cite{Ade:2015ava}. Upcoming large-scale structure (LSS) surveys have the potential to improve upon these limits significantly, through measurements of  scale-dependent bias \cite{Dalal:2007cu,Matarrese:2008nc,Afshordi:2008ru,Slosar:2008hx} and the bispectrum of biased tracers, such as galaxies~\cite{Scoccimarro:2003wn,Sefusatti:2007ih,Jeong:2009vd} (see \cite{Munoz:2015eqa,Baldauf:2016sjb,Tellarini:2016sgp,Yamauchi:2016wuc} for recent forecasts of particular shapes of PNG from the bispectrum of galaxies and 21-cm fluctuations).  The potential of LSS surveys stems from the fact that they provide a three-dimensional map of the universe compared to the two-dimensional map from CMB, and hence in principle they have access to larger number of modes. Extracting information from those modes, however, is intrinsically more challenging due to late-time nonlinearities. Obtaining significant improvement in constraints on PNG from LSS is contingent on our ability to model the observed statistics in the semi-nonlinear regime. The uncertainty in the theoretical model should be accounted for as a source of error. Additionally accurate estimates of the covariance matrix is essential.

Several previous studies, in particular in the context of quasi-single-field inflation, have forecasted the potential of future LSS surveys
to constrain the primordial non-Gaussianity due to massive scalar particles~\cite{Sefusatti:2012ye,Norena:2012yi,Gleyzes:2016tdh,Meerburg:2016zdz}. The detectability of massive spinning particles has been only recently studied in Refs.~\cite{MoradinezhadDizgah:2017szk,Bartolo:2017sbu}, albeit only through their effects on the galaxy bias. 
In this work, we extend these analyses by
studying how well galaxy-bispectrum measurements from next-generation spectroscopic surveys can observe, and identify, the non-Gaussianities arising from any extra particles with spins 2, 3, and 4.

To that goal, we develop an analytic expression of the bispectrum of primordial curvature perturbations due to the exchange of massive, spinning particles, which is currently only available in certain kinematical limits.
We present for the first time a new template for the primordial bispectrum due to massive particles with spins $s=2,3,4$ which matches the full numerical bispectrum to a good approximation. This template is valid for general triangle configurations, and accounts for both local (analytic) and non-local (non-analytic) effects contributing to the primordial bispectrum. 
We construct our template as the sum of these two contributions with a relative mass-dependent amplitude, set by extracting the analytical part of the full bispectrum due to a massive-particle exchange. 
Using this template, we make a forecast for how well two of the upcoming spectroscopic galaxy surveys,  
DESI~\cite{Aghamousa:2016zmz} and EUCLID~\cite{Laureijs:2011gra,Amendola:2016saw}, can constrain the masses and spins of these particles via measurement of the galaxy bispectrum.

We model the galaxy bispectrum in redshift space, at leading order in perturbation theory, accounting for contributions from both primordial non-Gaussianity and gravitational evolution \cite{Scoccimarro:1997st,Scoccimarro:2000sn,Sefusatti:2006pa,Scoccimarro:2003wn}. We assume a non-local Eulerian bias model \cite{McDonald:2009dh}, which for bispectrum at tree-level amounts to local linear and quadratic biases, and a non-local tidal bias. 
In modeling the redshift-space distortions (RSD), in addition to the linear Kaiser effect \cite{Kaiser:1987qv}, we also account for the non-linear smearing of the bispectrum due to the velocity dispersion of the tracers, i.e., the Finger-of-God (FOG) effect \cite{Jackson:2008yv}. Additionally, we also include the anisotropies induced by the Alcock-Paczynski (AP) effect \cite{Alcock}. We forecast constraints on the amplitude, and masses, of primordial non-Gaussianity from particles with spins $s= 2,3,4$, by marginalizing over the bias parameters, as well as the FOG velocity dispersion and the $\Lambda$CDM cosmological parameters.

We find that EUCLID and DESI perform similarly, and that for spinning particles with masses of the order of Hubble scale during inflation, we expect to detect any non-Gaussianities present above $f_{\rm NL} \approx 1$.  Moreover, we show that in the event of a detection of such primordial non-Gaussianity, we will be able to measure the mass of these particles with a relative uncertainty of the order of  $0.1/f_{\rm NL}$ for even-spin particles, and $1/f_{\rm NL}$ for odd spins. 
Ours is, to our knowledge, the first complete forecast that addresses quantitatively how galaxy bispectra can be used to detect---and characterize---massive particles with nonzero spin during inflation.

This paper is structured as follows. In Section~\ref{sec:Primordial}, we present the bispectrum template and discuss its qualitative features. In Section~\ref{sec:Galaxies}, we describe the galaxy bispectrum, which we employ to perform our forecasts. In section~\ref{sec:Fisher}, after describing our forecasting methodology and survey specifications, we present the forecasted constraints on PNG as well as cosmological parameters. We further discuss the impact of theoretical errors and non-Gaussian contributions to the bispectrum covariance on our forecast. We conclude in Section~\ref{sec:Conclusions}. In Appendix~\ref{app:template}, we detail the derivation of the bispectrum template. We further discuss the shape functions and their angular structure for the spinning particles in Appendix~\ref{sec:AppShapes}. Finally, in Appendix~\ref{sec:PowSp} we report the constraints on the non-Gaussian shapes considered from the galaxy power spectrum.

\section{Massive Spinning Particles During Inflation}
\label{sec:Primordial}
In this section, we will describe the imprints of massive particles with nonzero spin in the primordial bispectrum, and provide an ansatz for the shape of the bispectrum. We begin with some generalities about massive particles in effective theories in Section \ref{sec:3pt}. In Section \ref{sec:squeezed}, we describe the main qualitative features of the bispectrum due to massive spinning particles in the squeezed limit. We then present the bispectrum template in Section \ref{sec:angle} that we will implement in our Fisher forecasts in Section~\ref{sec:Fisher}. Finally, the shape of the bispectrum is discussed in Section \ref{sec:correlations}.

\subsection{Effective Description}\label{sec:3pt}
We begin by introducing the language that we use to describe interactions between massive particles and the inflationary perturbations. We then review and discuss the effects of massive spinning particles in the low-energy effective theory and their general properties during inflation, which will be helpful for understanding their main features in the three-point function. 

\paragraph{EFT of inflation.}
Inflation can be understood as a symmetry-breaking phenomenon. Due to a small departure from a de Sitter background, there exists a physical ``clock'' during inflation that reflects the time-dependent energy density of the inflationary background. This time dependence induces a preferred time foliation, thereby breaking time diffeomorphism invariance. This implies the existence of a Goldstone boson that encodes fluctuations along the direction of the broken symmetry, which we label by $\pi(t,\x)$. 
In spatially flat gauge, the Goldstone boson captures all the information contained in the cosmological perturbations in the scalar sector. This was the spirit of the effective field theory (EFT) of inflation developed in~\cite{Creminelli:2006xe, Cheung:2007st}, as a way of systematically characterizing the dynamics of inflationary perturbations without the need to specify the precise mechanism that drives the background evolution.

A more useful variable for describing cosmological correlation functions is the curvature perturbation $\zeta$ defined in comoving gauge, which is frozen on superhorizon scales. The relation between $\pi$ in spatially flat gauge and $\zeta$ in comoving gauge is given by
\begin{align}
	\zeta = -H\pi \, ,\label{eq:zetapi}
\end{align}
at linear order, while higher-order terms are slow-roll suppressed. Since the Hubble scale $H$ is nearly constant during inflation, the two fields are almost proportional to each other, allowing us to easily translate results from one gauge to the other. Below we will consider interactions between massive spinning degrees of freedom and the Goldstone boson $\pi$, and then compute the resulting correlation functions of $\zeta$.

\paragraph{Spinning operators.} 
Let us denote a totally symmetric, traceless massive spin-$s$ field by $\sigma_{\mu_1\cdots\mu_s}$. This field consists of $2s+1$ degrees of freedom with helicities that run from 0 to $s$. In order for this particle to contribute to the scalar three-point function at tree level, we will need a quadratic mixing between $\pi$ and $\sigma$. In this case, only the longitudinal mode of the spinning particle can contribute to the bispectrum. Kinematically, the most interesting effect arises when $\sigma$ carries the highest angular momentum, i.e.,~when $\sigma$ has the most number of spatial components. 
Let us therefore consider the following interaction Lagrangian between $\pi$ and $\sigma$~\cite{Lee:2016vti}:\footnote{More generally, there can also be contributions from interactions that are higher order in the spinning field, such as $\dot \pi \hat \sigma_{i_1...i_s}^2$ and $\hat \sigma_{i_1...i_s}^3$. These lead to diagrams that involve multiple exchanges of massive particles, but their leading non-local effect is the same as in the single-exchange case.} 
\begin{align}
	{\cal L}_{\rm int} = \lambda\hs\partial_{i_1\cdots i_s}\pi\hs\hat\sigma_{i_1\cdots i_s} + \tilde\lambda\hs\dot\pi\hs\partial_{i_1\cdots i_s}\pi\hs\hat\sigma_{i_1\cdots i_s}\, ,\label{Lint}
\end{align}
where $\lambda$ and $\tilde\lambda$ are coupling constants, $\partial_{i_1\cdots i_s}\equiv \partial_{i_1}\cdots\partial_{i_s}$, and hatted symbols indicate the traceless part. 

We are interested in the effect of the above interactions on the dynamics of $\pi$. Note that the above mixing interactions modify the linearized, on-shell equation of motion for the spinning field, schematically to the form
\begin{align}
	\big(\Box - m^2\big)\hat\sigma_{i_1\cdots i_s} = -\lambda\hs\hat\partial_{i_1\cdots i_s}\pi\, ,
\end{align}
where $\Box\equiv \nabla^\mu\nabla_\mu$. One should note that the presence of the traceless field on the left-hand side forces the derivative structure on the right-hand side to be traceless as well. Classically, integrating out $\sigma$ amounts to substituting the solution of the equation of motion back to the Lagrangian. If the mass $m$ of the particle is sufficiently large compared to the typical energy scale of the experiment (in our case, the Hubble scale $H$), then we can expand the resulting Lagrangian in inverse powers of $m^2$, and only keep the leading term in the expansion. This corresponds to the following replacement rule: 
\begin{align}
	\hat\sigma_{i_1\cdots i_s} \, \xrightarrow{m\hs\gg\hs H} \, \frac{\lambda}{m^2}\hat\partial_{i_1\cdots i_s}\pi+\cdots\, ,\label{replace}
\end{align}
where the ellipses denote higher-derivative corrections suppressed by ${\cal O}(\Box/m^2)$ relative to the leading term. 

Let us further examine the consequences of integrating out a massive spinning degree of freedom on the low-energy dynamics. 
The self-interaction of $\pi$ that results from integrating out a massive spin-$s$ field takes the form
\begin{align}
	{\cal L}_{\rm \pi} = \rho\hs \dot\pi(\hat\partial_{i_1\cdots i_s}\pi)^2\, ,\label{Lpi}
\end{align}
where $\rho \equiv \lambda\tilde\lambda/m^2$ is an effective coupling parameter. We wish to highlight several features of this self-coupling. First, we see that the information about the mass of the particle is lost once it is integrated out, since the mass dependence gets absorbed into the overall coefficient and becomes degenerate with the coupling constant. This is the usual feature of effective field theories that ultraviolet (UV) physics becomes decoupled from infrared (IR) dynamics. In contrast, the resulting local operator still carries the spin dependence of the integrated-out particle, as it retains the traceless combination of spatial derivatives, mimicking the polarization of the original particle. In general, individual higher-dimensional operators in the effective theory can carry arbitrary coefficients, which are to be constrained by experiments. 
Seeing the presence of this operator with such a particular derivative structure in a low-energy effective theory would then hint at the presence of a massive spinning particle in the ultraviolet theory. Having said that, this operator is accompanied by more derivatives than the original spinning operator. We will examine the phenomenological difference between the UV and IR spinning operators in \S\ref{sec:correlations}.

\paragraph{Local vs.~non-local.}
There are two distinguishing effects that heavy particles can lead to on the dynamics of low-energy effective theories, which we will refer to as local and non-local effects, following Ref.~\cite{Arkani-Hamed:2015bza}. Local effects can be fully described by effective Lagrangians involving only light degrees of freedom with heavy particles integrated out, cf.~\eqref{Lpi}.
On the other hand, the effects of physically created heavy particles are not captured by an effective field theory expansion. Instead, to characterize these effects one needs the full description of interactions in terms of the heavy field, cf.~\eqref{Lint}. 

These two effects can be understood by examining the free propagation of particles in de Sitter space. The distinctive properties of massive particles during inflation become pronounced when their wavelengths are stretched outside the horizon at late times. Their behavior in this regime is dictated purely by the symmetries of the cosmological background. For instance, the dilatation (scaling) symmetry implies that a massive spin-$s$ particle behaves in the late-time limit as
\begin{align}
	\sigma_{i_1\cdots i_s}(\eta\to 0,\x) = \sum_\pm (-\eta)^{\Delta_\pm-s}\sigma^\pm_{i_1\cdots i_s}(\x)\, ,
\end{align}
where $\eta$ is conformal time and\footnote{We deviate from the standard notation and use $\nu$ to denote the effective mass parameter in order to avoid potential confusion with the usage of the notation $\mu$ as an angular variable in later sections.}
\begin{align} \label{eq:nu}
	\Delta_\pm  \equiv \frac{3}{2} \pm i\nu \, ,\quad 
	\nu \equiv  \sqrt{\frac{m^2}{H^2}-\left(s-\frac{1}{2}\right)^2}\, ,
\end{align}
are the scaling dimensions of the field, fixed by the particle's mass for a given spin. The temporal and spatial dependence of the field hence factorizes at late times. In this limit, the isometries of de Sitter space act as a three-dimensional conformal group on the spatial future boundary. The form of the spatial dependence $\sigma^\pm_{i_1\cdots i_s}$ projected on this slice is then fully fixed by the conformal symmetry. For light particles with an imaginary $\nu$, the late-time behavior is dominated by one of the modes, decaying monotonically in time. In contrast, for massive particles with a real $\nu$, the two modes become complex conjugates to each other, and the mode oscillates in time. 

Let us be more explicit and elaborate on the qualitative features. We will restrict to the case of real $\nu> 0$ in our subsequent discussions. Introducing spatially null vectors $\eps=(\cos\alpha,\sin\alpha,i)$ and $\tilde\eps=(\cos\beta,\sin\beta,-i)$, the late-time two-point function of a massive spin-$s$ particle in momentum space is given by~\cite{Arkani-Hamed:2015bza, Lee:2016vti}
\begin{align}
	\lim_{\eta,\tilde\eta\to 0}\langle \varepsilon\cdot\sigma(\eta)\hs \tilde\varepsilon\cdot\sigma(\tilde\eta)\rangle' &= \frac{(H^2\eta\tilde\eta)^{3/2-s}}{4\pi H}\sum_{\lambda=-s}^s e^{\lambda\chi} \frac{(2s-1)!!\hs s!}{(s-\lambda)!(s+\lambda)!} \Big[G_{\rm L}(\eta,\tilde\eta)+G_{\rm NL}(\eta,\tilde\eta)\Big]\, ,
	\end{align}
where $\varepsilon\cdot\sigma\equiv \varepsilon_{i_1}\cdots \varepsilon_{i_s}\sigma_{i_1\cdots i_s}$ and $\chi=\alpha-\beta$. We defined
\begin{align}
	G_{\rm L}(\eta,\tilde\eta) &=\frac{\pi}{\nu\sinh\pi\nu}\left[e^{\pi\nu}\left(\frac{\eta}{\tilde\eta}\right)^{i\nu}+e^{-\pi\nu}\left(\frac{\eta}{\tilde\eta}\right)^{-i\nu}\right] ,\\[4pt]
	G_{\rm NL}(\eta,\tilde\eta) &=\frac{\Gamma(\frac{1}{2}+s-i\nu)\Gamma(\frac{1}{2}+\lambda+i\nu)}{\Gamma(\frac{1}{2}+s+i\nu)\Gamma(\frac{1}{2}+\lambda-i\nu)}\left(\frac{k^2\eta\tilde\eta}{4}\right)^{i\nu}\Gamma(-i\nu)^2+c.c.\, ,\, 
\end{align}
which refer to the local and non-local terms, respectively. Let us make some remarks on the form of this two-point function. First, the overall factor $(\eta\eta')^{3/2}$ reflects the volume dilution of the particle number density with the expansion of the universe.\footnote{The additional dependence $(H^2\eta\eta')^{-s}$ is due to some extra scale factors that are cancelled when the spatial indices are properly contracted with the metric.} The helicities $\lambda$ of the massive particle are summed over, each depending on a particular phase with some purely kinematical factor. We also see some stark differences between the local and non-local terms, which we explain below:
\begin{itemize}
	\item The overall mass-dependent factor of the local term behaves as $1/\nu$ in the limit $\nu\to\infty$. This corresponds to the leading effect that would be captured by an effective field theory expansion. We also see that the the two time arguments in $G_{\rm L}(\eta,\tilde\eta)$ precisely cancel each other in the equal-time limit, so that its overall time dependence is simply given by the volume dilution. Moreover, there is no dependence on $k$, which implies that it only describes correlations at coincident points in position space. This local contribution is not fixed by the conformal symmetry, and all helicities are treated on an equal footing as in flat space. 
	\item Instead of a power-law suppression, the non-local part behaves as $e^{-\pi\nu}\sim e^{-\pi m/H}$ in the large-mass limit, meaning that its size only becomes appreciable when $m\sim H$. This Boltzmann-like factor indicates that the particle is physically produced in the time-dependent background, which effectively behaves as a thermal bath with temperature $T=H/2\pi$. Unlike the local term, the two time arguments do not cancel in the equal-time limit, giving an oscillatory behavior in $\ln(k\eta)$ with a frequency $\nu$.\footnote{The fact that the oscillations are logarithmic in conformal time is due to the fact that the space is expanding exponentially. In terms of physical time $t$, the oscillations take the usual form $e^{\pm imt}$ for large masses.} There is also a non-analytic dependence on $k$, which, again, represents an effect that is not captured by an effective field theory expansion, and corresponds long-distance correlations in position space. The amplitude of each helicity component belonging to the non-local part is precisely fixed by the conformal symmetry. Measuring the consistency between different helicities would be a non-trivial check that these particles were indeed produced in an inflationary, quasi-de Sitter background.
\end{itemize}
The presence of the non-local term highlights a major difference between conventional particle colliders and the cosmological collider. In order to physically create particles and observe the corresponding resonance in the former case, the center of mass energy of the collisional process must be higher than the particle's mass. In contrast, any particles that are present during inflation are physically created by the time dependence of the background, albeit with a rate of production that is exponentially suppressed for large masses. This leaves a window of direct detection of heavy particles during inflation in the mass range $m \lesssim {\cal O}(sH)$.

\subsection{The Squeezed Limit}\label{sec:squeezed}
We will consider the leading imprint of massive particles with spin on the primordial bispectrum from a tree-level exchange. Using the standard in-in formalism, it can be shown that the interactions between $\pi$ and $\sigma$ in Eq. \eqref{Lint} generate a primordial bispectrum of the form
\begin{align}
\langle\zeta_{\k_1}\zeta_{\k_2}\zeta_{\k_3}\rangle'\propto P_s(\hat\k_1\cdot\hat\k_3)\, {\cal I}^{(s)}(k_1,k_2,k_3) + \text{5 perms}\, ,
\end{align}
where $P_s$ is the Legendre polynomial, hatted vectors denote unit vectors, $k_i\equiv |\k_i|$, and the prime on the expectation value indicates the momentum-conserving delta function has been stripped off. The function ${\cal I}^{(s)}(k_1,k_2,k_3)$ carries the momentum dependence, whose integral representation can be found in Appendix~\ref{app:template}. Before describing the shape of the total bispectrum, we will find it instructive to look at the shape that arises from each permutation separately. Different channels give rise to complicated momentum dependencies in general triangular configurations, but they become dramatically simplified in a special kinematical limit known as the squeezed limit, in which one side of the triangle formed by momentum vectors is taken to be small. Taking the limit $k_1\ll k_2\approx k_3$, the composition of the bispectrum from different permutations can be diagrammatically expressed as 
\begin{align}
\lim_{k_1\ll k_3}\langle\zeta_{\k_1}\zeta_{\k_2}\zeta_{\k_3}\rangle' \ \propto\ \underbrace{\includegraphicsbox[scale=.45]{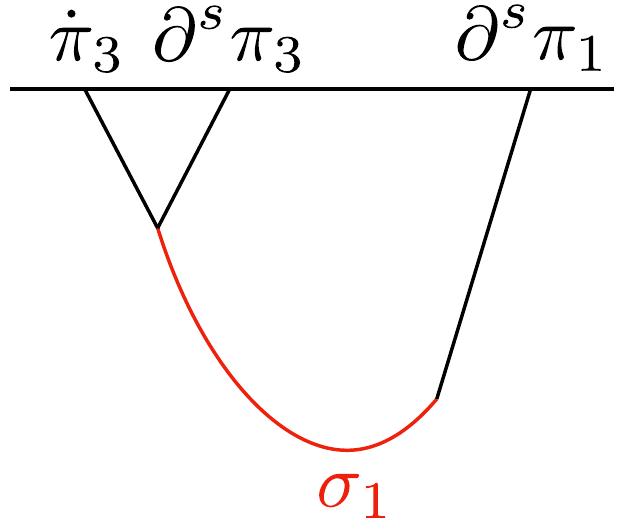}}_{I_1\, \equiv\, I^{(s)}(k_1,k_3,k_3)}\ +\ \underbrace{\includegraphicsbox[scale=.45]{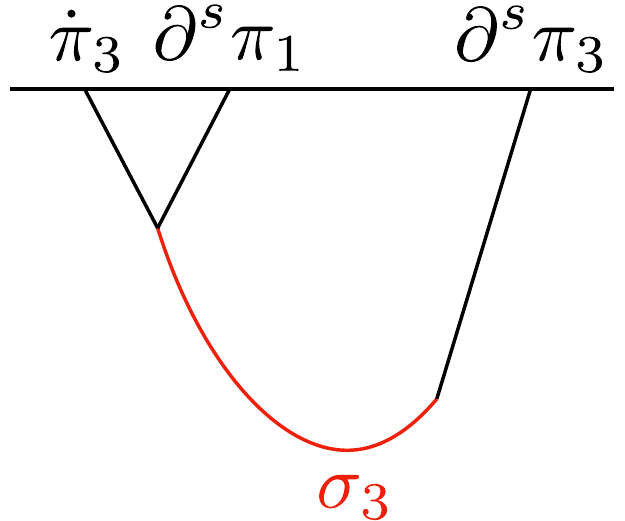}}_{I_2\, \equiv\, I^{(s)}(k_3,k_1,k_3)} \ +\ \underbrace{\includegraphicsbox[scale=.45]{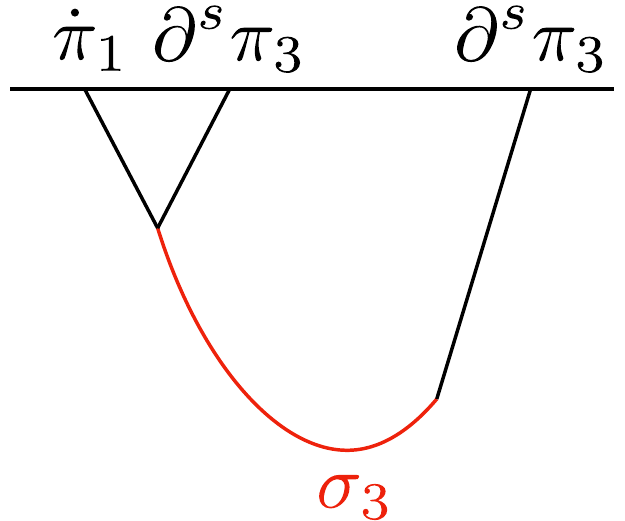}}_{I_3\, \equiv\, I^{(s)}(k_3,k_3,k_1)}  \ ,\label{eq:diagram}
\end{align}
where $\{\pi_n,\sigma_n\}\equiv \{\pi_{\k_n},\sigma_{\k_n}\}$ and $I^{(s)}(k_1,k_2,k_3)\equiv  P_s(\hat\k_1\cdot\hat\k_3)\,{\cal I}^{(s)}(k_1,k_2,k_3)$. Let us describe some qualitative features of each of these diagrams:
\begin{itemize}
	\item $I_1$: In this channel, the intermediate particle directly mixes with the soft external mode $k_1$, and hence also carries a soft momentum. The long-wavelength massive mode starts propagating on superhorizon scales when $|k_1\eta|\sim 1$ until decaying into two $\pi$ fluctuations when $|k_3\eta|\sim 1$. The oscillatory time evolution of the massive particle is then transcribed into oscillations in $\ln (k_1/k_3)$ in the bispectrum, while the volume dilution leads to a suppression $(k_1/k_3)^{3/2}$. The cubic vertex involves the contraction between hard external momenta and the soft longitudinal polarization tensor, generating an angular dependence $P_s(\hat\k_1\cdot\hat\k_3)$. This channel is the most interesting one, as it contains both the mass and spin dependence of the intermediate particle.
	\item $I_2$: Here, the spatial gradient mode of the cubic vertex is taken to be the soft mode. The intermediate particle propagates over a very short distance on subhorizon scales, and the tree exchange essentially collapses into a contact diagram. The overall momentum dependence will be analytic in $k_1$, and hence no oscillations will be observed in this channel. On the other hand, we still have a non-trivial momentum contraction involving the polarization tensor, which in fact gives the same angular dependence as the diagram $I_1$. The difference from $I_1$, however, is that now the external leg carries a soft momentum, which gives an extra suppression $(k_1/k_3)^s$ in the squeezed limit. 
	\item $I_3$: This channel corresponds to taking the soft limit of the $\dot\pi$ mode in the cubic vertex. The time derivative will lead to the suppression factor of $(k_1/k_3)^2$ in the squeezed limit, which is less severe than that of the diagram $I_2$ for generic spins, but notice that this diagram contains neither the oscillations nor the angular dependence. It can therefore be viewed as an ``effective noise'' channel that doesn't contain any information on the nature of the extra particle.
\end{itemize}
As we explained previously, there will be two types of contributions to the bispectrum from massive particles which arise from local and non-local effects. We will refer to the shapes that arise from local and non-local effects as {\it analytic} and {\it non-analytic} shapes, respectively, for reasons explained below.

\paragraph{Analytic shape.}
The scaling behavior of the bispectrum in the squeezed limit due to local effects simply follows from the derivative structure of the local operator in Eq.~\eqref{Lpi}. In the squeezed limit, $k_1\ll k_2\approx k_3$, the analytic part behaves as
\begin{align}
	\lim_{k_1\ll k_3}\langle\zeta_{\k_1}\zeta_{\k_2}\zeta_{\k_3}\rangle ' \propto \frac{1}{k_1^3k_3^3}\left[\left(\frac{k_1}{k_3}\right)^2+\cdots + \left(\frac{k_1}{k_3}\right)^sP_s(\hat \k_1\cdot\hat \k_3)\right] + (\k_2\leftrightarrow\k_3)\, .
\end{align}
The leading piece with the scaling $(k_1/k_3)^2$ corresponds to taking the soft limit of the external leg $\dot\pi$ or the diagram $I_3$ in Eq.~\eqref{eq:diagram}. Notice that, as pointed out previously, this particular permutation doesn't induce any angular dependence. The Legendre behavior arises in the other channel $I_3$ for which we take the soft $\hat\partial_{i_1\cdots i_s}\pi$ leg, whose number of derivatives makes it to appear at order $(k_1/k_3)^s$. This means that the angular dependence will only appear at leading order for spin 2, while higher spin dependence will be subleading in the squeezed limit. Observing the imprints of higher-spin ($s>2$) particles from the analytic shape hence requires analyzing the angular dependence in general triangular configurations.

Note that odd-spin Legendre polynomials gain an extra suppression in $k_1/k_3$ when summed over permutations due to the momentum conservation, whereas even-spin terms are symmetric under momentum exchange. This means that only even powers of the ratio $k_1/k_2$ can appear in the squeezed-limit expansion. This analytic momentum dependence has a clear meaning when we go to position space. Because the analytic shape, by definition, arises from local interactions, it cannot induce correlations between two distant positions on a fixed time slice. When we Fourier transform the position-space correlator, this locality is reflected in an analytic dependence on momenta~\cite{Maldacena:2011nz,Arkani-Hamed:2015bza}.

\paragraph{Non-analytic shape.}
The non-local effects due to the massive intermediate particle only make an appearance in the three-point function when the internal mode propagates over a long distance or carries a soft momentum, i.e.,~in the diagram $I_1$. In the squeezed limit this diagram behaves as
\begin{align}
	\lim_{k_1\ll k_3}\langle\zeta_{\k_1}\zeta_{\k_2}\zeta_{\k_3}\rangle ' \propto \frac{1}{k_1^3k_3^3}\left(\frac{k_1}{k_3}\right)^{3/2}\cos\left[\nu\ln\frac{k_1}{k_3}+\varphi\right] P_s(\hat \k_1\cdot\hat \k_3)+ (\k_2\leftrightarrow\k_3)\ .\label{nonanalytic}
\end{align}
For $m>(s-1/2)H$, the parameter $\nu$ becomes real and the bispectrum oscillates logarithmically in the momentum ratio. Recall that the Maldacena's consistency relation~\cite{Maldacena:2002vr,Creminelli:2004yq} states that any physical effects in single-field inflation appear at order $(k_1/k_3)^2$~\cite{Creminelli:2012ed,Hinterbichler:2013dpa,Creminelli:2013cga}. The squeezed bispectrum therefore provides a clean detection channel for extra particles during inflation, which produce a characteristic non-analytic momentum dependence $(k_1/k_3)^{3/2}$. For odd spins, the scaling instead becomes $(k_1/k_3)^{5/2}$ after summing over the permutations.

A few comments are in order regarding cases where particles are lighter, corresponding to cases with an imaginary $\nu$. When this is the case, the bispectrum no longer becomes oscillatory, and can have scalings that are less suppressed than $(k_1/k_3)^{3/2}$. Having said that, the unitarity mass bound in de Sitter space, $m^2\ge s(s-1)H^2$~\cite{Higuchi:1986py,Deser:2001us}, sets the smallest allowed mass value for massive particles with $s\ge 2$ during inflation. When the particle's mass saturates this bound, the suppression becomes linear in $k_1/k_3$. Interestingly, at some discrete mass points below the unitarity bound, particles can gain gauge symmetries which results in a less propagating degrees of freedom than the naive massive case, which is a phenomenon called ``partial masslessness''~\cite{Deser:1983mm, Deser:2001xr}. Being lighter, some of these partially massless particles can contribute to soft limits of correlation functions without any suppression, making their prospective detection easier~\cite{Baumann:2017jvh,Franciolini:2017ktv}. In this paper, we will focus on the effects of massive particles with real values of $\nu$, leaving the investigation of a more general particle spectrum to future work.

\subsection{Bispectrum Template}\label{sec:angle}

As we saw above, while the non-analytic shape constitutes a clear signature of new particles, its associated amplitude becomes exponentially suppressed for $m\gg H$. 
The information on the particle mass is thus lost in the large-mass limit, whereas its spin dependence is still partially contained in the analytic shape.
In general, detecting non-Gaussianity from local operators would likely happen first, and only then could one look for the actual particle creation by examining the scaling behavior of the squeezed bispectrum. Analyzing this likelihood more quantitatively requires comparing the relative sizes of the analytic and non-analytic shapes.

For these reasons, we will take both the analytic and non-analytic shapes into account in building our bispectrum template. A closed-form expression for the bispectrum due to a tree-level particle exchange in arbitrary triangular configurations is not currently available in the literature, due to the difficulty of performing the integrations involved in computing the correlator. Rather than pursuing a numerical approach, we will introduce an ansatz for the shape function that agrees with the numerical bispectrum to a sufficiently good approximation. Essentially, it is given by a linear combination of the analytic and non-analytic shapes, with some relative amplitude. The basic reason why this works is because, as we discussed earlier, the exchange diagram approaches a contact diagram away from the squeezed limit, in which case the shape is mostly dictated by the analytic part.

Leaving the details to Appendix~\ref{app:template}, we will consider the following bispectrum template: 
\begin{align}
	B^{\rm spin}_\nu(k_1,k_2,k_3) =  \alpha^{(s)}\hskip -1pt f_{\rm NL}^{\rm spin}\big[B^{\small\rm A}(k_1,k_2,k_3)+ B^{\text{NA}}_\nu(k_1,k_2,k_3)\big]\, ,
	\label{eq:Bfull}
\end{align}
where
\begin{align}
	B^{\rm A}(k_1,k_2,k_3) & = \frac{\hskip -0.5pt P_s(\hat \k_1\cdot\hat \k_3)}{k_2(k_1k_3)^{3-s}k_t^{2s+1}}\Big[(2s-1)\big((k_1+k_3)k_t+2sk_1k_3\big)+k_t^2 \Big] + \text{5 perms}\, ,\label{Bangle}\\[4pt]
	B^{\text{NA}}_\nu(k_1,k_2,k_3) & = \frac{r^{(s)}\hskip -0.5pt(\nu)}{k_1^3k_3^3}\left(\frac{k_1}{k_3}\right)^{3/2}\cos\left[\nu\ln\frac{k_1}{k_3}+\varphi\right]\!P_s(\hat\k_1\cdot\hat\k_3) \Theta(x_*k_3-k_1)+ \text{5 perms}\, ,\label{BNA}
\end{align}
are the shapes for the analytic and non-analytic parts, respectively, $k_t\equiv k_1+k_2+k_3$, and $\Theta$ is the Heaviside step function. The mass-dependent phase $\varphi$ is computable for a given type of interaction, and $r^{(s)}$ is the mass-dependent relative amplitude between the analytic and non-analytic parts for a given spin. 
Since the non-analytic shape is only valid in the squeezed limit, we introduce a cutoff $x_*=0.1$ for the range of the momentum ratio for each permutation. We use the standard measure of the size of the bispectrum by the nonlinearity parameter
\begin{align}
	\fnl \equiv \frac{5}{18}\frac{B_\zeta(k,k,k)}{P_\zeta^2(k)}\, ,
\end{align}
defined as its amplitude in the equilateral configuration, normalized with respect to two power spectra.  
A spin-dependent coefficient $\alpha^{(s)}$ is inserted in Eq.~\eqref{Bangle} in such a way that $f_{\rm NL}^{\rm spin}=\frac{5}{18} B^{\rm A}(1,1,1)/P_\zeta^2(1)$. 

\begin{figure}[t!]
	\centering
	\includegraphics[scale=0.75]{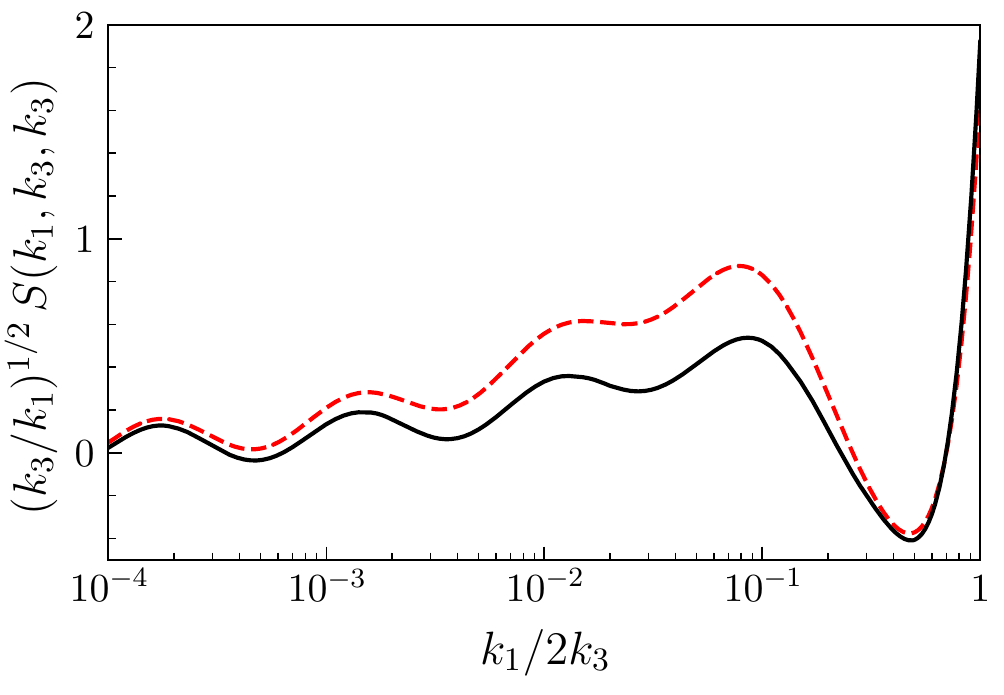}\quad\includegraphics[scale=0.75]{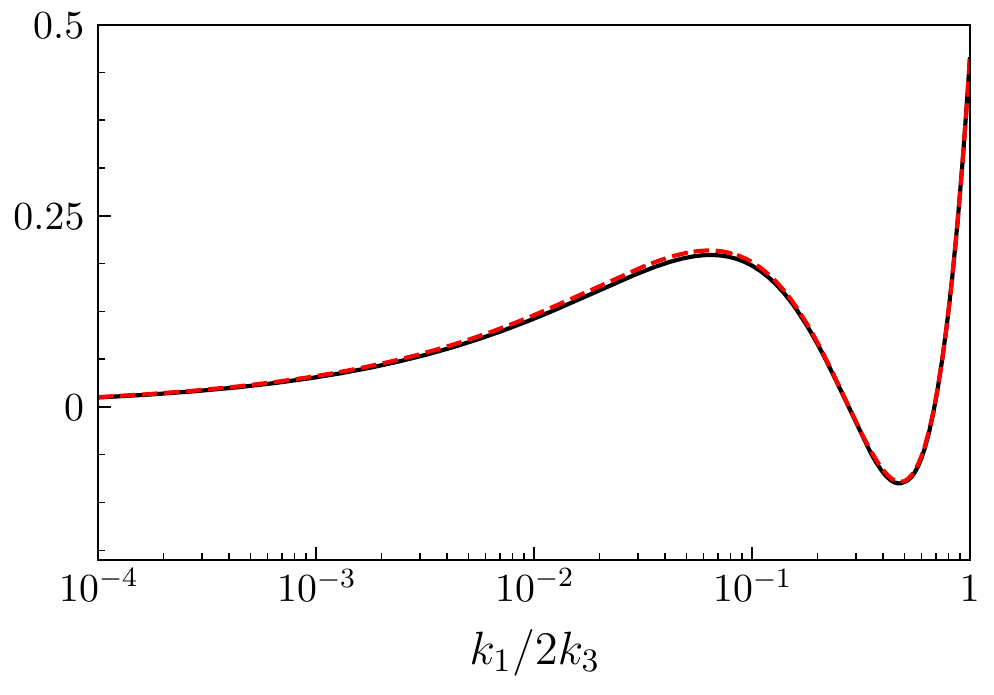}
	\caption{Bispectrum due to the exchange of a massive spin-2 particle as function of the momentum ratio for $\nu=3$ (\emph{left}) and $\nu=6$ (\emph{right}). The solid and dashed lines correspond to the numerical result and analytical template, respectively.}\label{fig:template}
\vspace{-.1in}
\end{figure}

The analytic part of the template is the shape that arises from the self-interaction in Eq.~\eqref{Lpi}. We expect this to account for the dominant shape of the bispectrum in the large-mass limit. An important feature of our template is the relative amplitude $r^{(s)}$, predetermined for each spin. To fix the ratio, we first compute the amplitude of the analytic part of the full tree-exchange bispectrum in a squeezed configuration, and then use it to fix the overall amplitude of the analytic shape in Eq.~\eqref{Bangle}. This provides an analytic formula for the full bispectrum shape for generic mass values beyond the squeezed limit, which is easy to implement in data analysis. The derivation of the template can be found in Appendix~\ref{app:template}.

Figure~\ref{fig:template} shows the fitting of our template \eqref{eq:Bfull} to the numerically computed bispectrum for $s=2$ and two mass values corresponding to $\nu=\{3,6\}$ in terms of the dimensionless shape function defined in Eq.~\eqref{eq:shape}. We see that the template provides a very good fit in the high-mass regime, where it becomes well approximated purely by the analytic shape. In contrast, the fitting is not as perfect in the low-mass regime. On the one hand, the amplitude of the oscillations, as well as the non-squeezed shapes of the numerical bispectrum and the template, precisely agree with each other. On the other hand, there is a small offset in the overall amplitude in the intermediate range of momentum ratios. As we explain below, this is a reflection of the fact that we cannot fully drop higher-derivative corrections when the mass of the integrated-out particle is sufficiently close to $H$. Nevertheless, the characteristic oscillatory behavior of the bispectrum in the low-mass regime is still well captured by the template, which suffices to estimate how well the masses of these particles can be measured.

\begin{figure}[t!]
	\centering
	\includegraphics[scale=0.75]{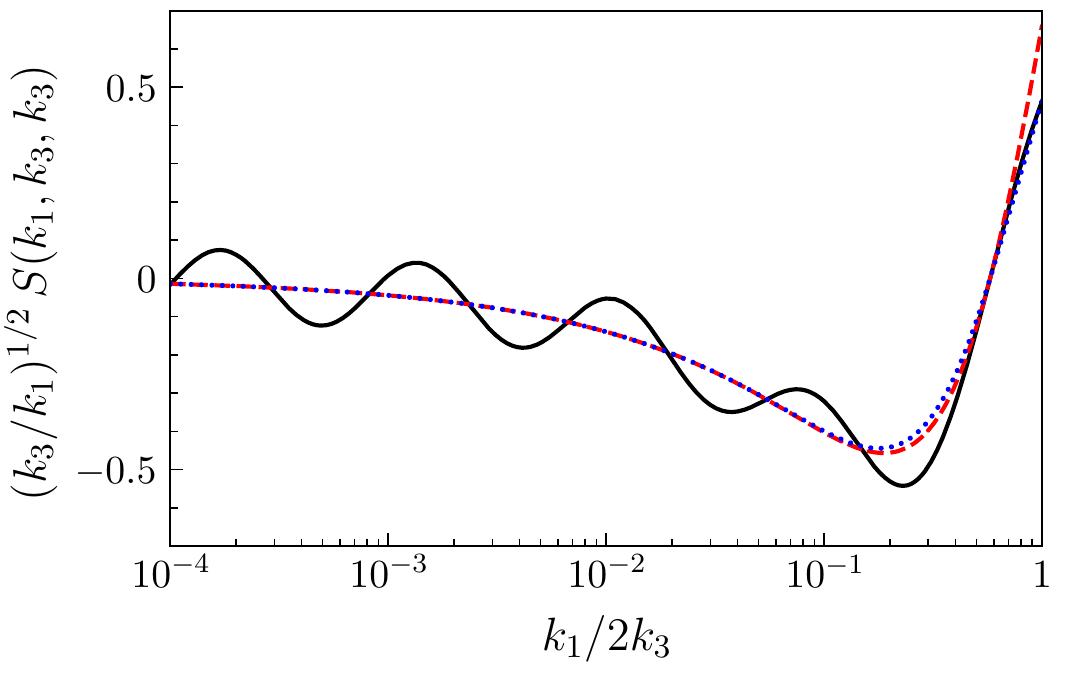}\quad\includegraphics[scale=0.75]{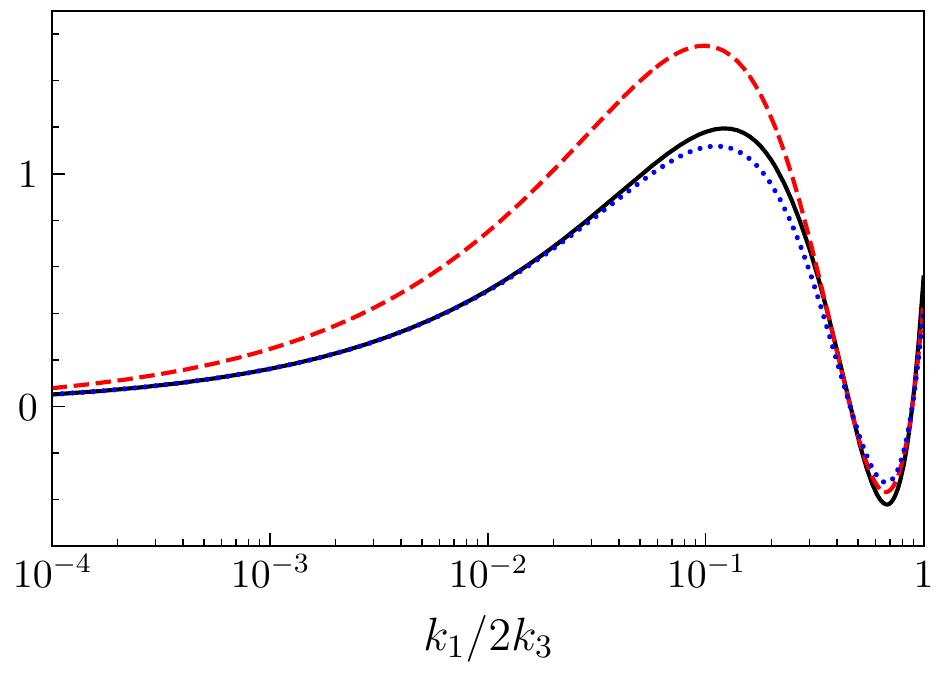}
	\caption{Bispectrum due to the exchange of a massive spin-2 particle as a function of the momentum ratio for different permutations. The solid (black), dotted (blue), and dashed (red) lines correspond to the numerical result, analytical template with and without the higher-derivative terms, respectively. The left and right plots correspond to the diagrams $I_1$ and $I_3$ in Eq.~\eqref{eq:diagram}, respectively. }\label{fig:template2}
	\vspace{-.1in}
\end{figure}

Let us further examine the source of the discrepancy by decomposing the full bispectrum into different channels. Figure~\ref{fig:template2} shows a comparison between the numerical result and the template for two permutations of the bispectrum that correspond to the diagrams $I_1$ and $I_3$ in Eq.~\eqref{eq:diagram}. The reason why the fitting remains accurate for the left plot of Fig.~\ref{fig:template2} is because we fix the amplitude of the analytic shape precisely with respect to the diagram that contains the oscillations. The overall inaccuracy of the fitting can be traced to the fact that the field $\hat\sigma_{ij}$ cannot be replaced by $\hat\partial_{ij}\pi$ alone for small $\nu$, in which case one needs to keep track of higher-derivative corrections in Eq.~\eqref{replace}. These corrections involve terms such as $\dot\pi\partial_k^2(\hat\partial_{ij}\pi)^2$, which lead to more general contact diagrams. Notice that this particular correction is only relevant in the channel with a soft $\dot\pi$ mode, i.e.,~the diagram $I_3$, and produces a negligible effect in the other channels due to the extra spatial gradients. The template including this higher-derivative correction is indicated by the blue dotted lines in Fig.~\ref{fig:template2}, and we see that it helps to fit the numerical bispectrum better. We thus expect that the proper inclusion of the next-to-leading-order terms can lead to a more accurate template for small $\nu$. Nevertheless, these corrections only modify the ``effective noise'' channel and hence do not alter our ability to constrain the mass of the particle. For simplicity, we will therefore use the template Eq.~\eqref{eq:Bfull} for our Fisher analysis in Section~\ref{sec:Fisher}.

In the rest of the paper, we will also refer to three other commonly-studied types of bispectra: the local~\cite{Gangui:1993tt,Wang:1999vf,Verde:1999ij,Komatsu:2001rj}, equilateral~\cite{Babich:2004gb,Creminelli:2005hu}, and quasi-single-field~\cite{Chen:2009zp} non-Gaussianity. For completeness, we list their bispectrum templates below:\footnote{We have stated scale-invariant formulas here for simplicity. For our Fisher analysis in Section~\ref{sec:Fisher}, we reintroduce the small scale dependence through the replacement $k_i\to k_i^{1+(1-n_s)/3}$ for $i=1,2,3$.}

\begin{subequations}
\begin{align}
	B^{\rm loc}(k_1,k_2,k_3) &=  \frac{6}{5} A_\zeta^2 f_{\rm NL}^{\rm loc}\,\frac{k_1^3+k_2^3+k_3^3}{(k_1k_2k_3)^3}\, ,\label{Blocal}\\
	B^{\rm eq}(k_1,k_2,k_3) &=   \frac{18}{5} A_\zeta^2 f_{\rm NL}^{\rm eq}\,\frac{(k_1+k_2-k_3)(k_2+k_3-k_1)(k_3+k_1-k_2)}{(k_1k_2k_3)^3} \, , \label{Beq}\\
	B_{\tilde\nu}^{\rm qsf}(k_1,k_2,k_3) &=  \frac{54\sqrt{3}}{5} A_\zeta^2  f_{\rm NL}^{\rm qsf}\,\frac{k_1k_2k_3}{\sqrt{k_t}}\frac{Y_{\tilde\nu}(8\kappa)}{Y_{\tilde\nu}(8/27)}\, ,\label{BQSF}
\end{align}
\end{subequations}
where $\kappa\equiv k_1k_2k_3/k_t^3$, $Y_{\tilde\nu}$ is the Bessel function of the second kind of degree $\tilde\nu \equiv \sqrt{9/4-m^2/H^2}$, and  $ A_\zeta  = k^3 P_\zeta(k ) $ is the amplitude of the power spectrum of primordial curvature fluctuations ($A_\zeta =  2 \pi^2 A_s$ with $A_s = (2.142 \pm 0.049) \times 10^{-9}$ from Planck data \cite{Ade:2015lrj}).

\subsection{Shape Functions}\label{sec:correlations}

The bispectrum of our consideration is (nearly) scale invariant, in which case we can factor out the overall $k^{-6}$ scaling, so that the shape can be described by a function of two variables. It is then convenient to introduce the dimensionless shape function defined by
\begin{align}
	S(k_1,k_2,k_3)\equiv \frac{(k_1k_2k_3)^2}{A_s^2}B(k_1,k_2,k_3)\, . \label{eq:shape}
\end{align}
In order to see the angular dependence, we fix the ratio between two momenta and vary the angle between them. Defining the momentum ratio $r\equiv k_1/k_2$ and the angle $\cos\theta\equiv\hat\k_1\cdot\hat\k_2$, the shape function as a function of $\{r,\theta\}$ is given by
\begin{align}
	{\cal S}(r,\theta)\equiv S\big(1,r,\sqrt{1+r^2+2\hs r\cos\theta}\big)\, .
\end{align}

\begin{figure}[t!]
	\centering
	\includegraphics[scale=0.75]{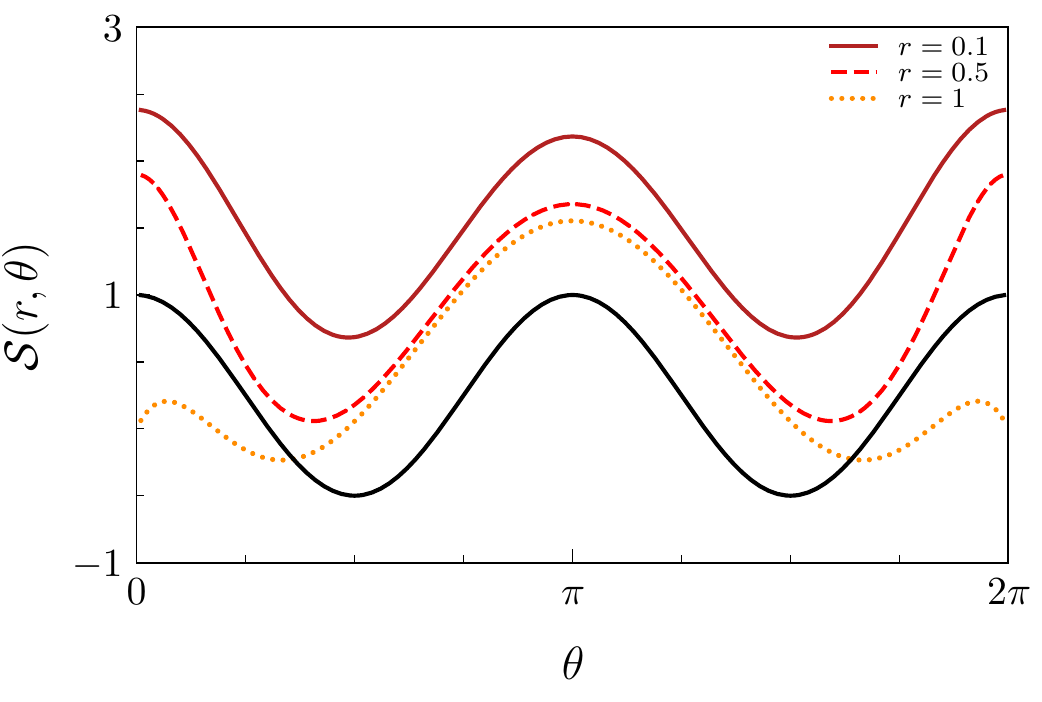}\ \includegraphics[scale=0.75]{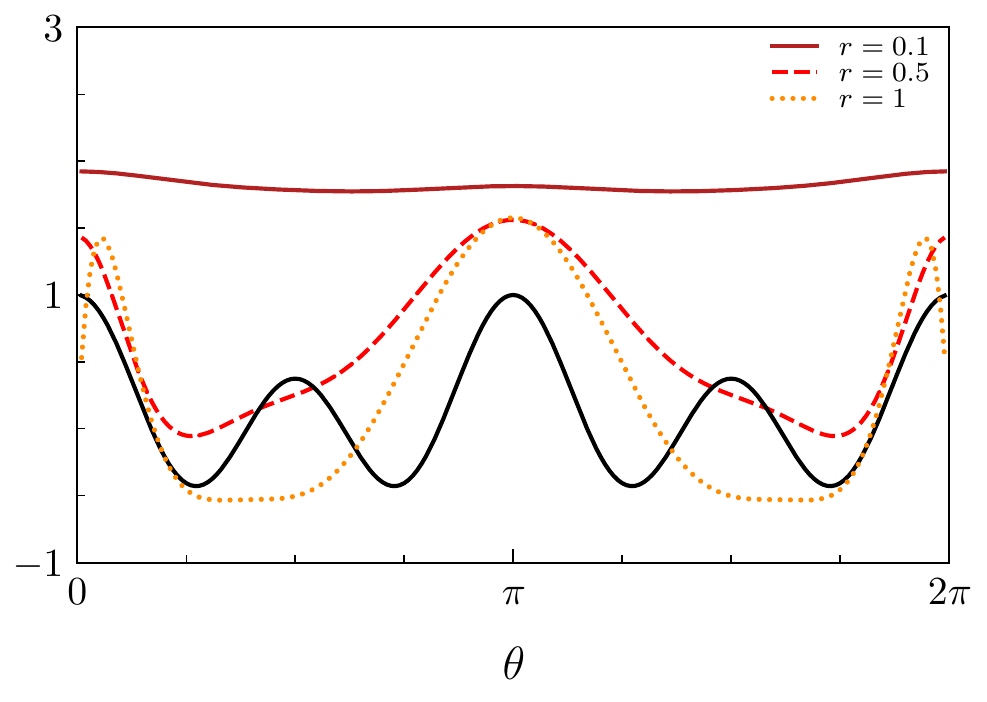}
	\caption{Analytic shape functions for spin 2 (\emph{left}) and spin 4 (\emph{right}) as a function of the base angle $\theta=\cos^{-1}(\hat \k_1\cdot \hat \k_3)$ for fixed ratios of $r=k_1/k_2$. The black curve shows the pure Legendre behavior, whereas the solid, dashed, and dotted colored curves represent the angular dependence when $r= 0.1$, 0.5, and 1, respectively. For easier comparison, each curve has been normalized with respect to the Legendre polynomials. \label{fig:polar}
}
\end{figure}

Figure~\ref{fig:polar} shows the angular dependence of the shape function for the analytic part for $s=\{2,4\}$ for a range of momentum configurations fixing two sides of the triangle. We see some crucial differences between these two cases with different spins. For spin $2$, the angular dependence converges to the pure Legendre behavior as the triangle becomes squeezed. On the other hand, the angular dependence completely disappears in the same limit for spin $4$. As we explained previously, this is due to the extra derivative suppression of the local operator as the spin increases. We see that there is a non-trivial angular dependence at the configuration with $r=1$, which, however, does not quite match the pure Legendre spin-4 behavior. This suggests that it would be difficult in practice to constrain the spin dependence by just looking at the angular dependence of the analytic shape beyond spin $2$, unless there is enough signal-to-noise away from the squeezed limit. This provides another motivation to consider the full bispectrum template including the non-analytic part, which carries both the spin and mass dependences. 

\paragraph{Shape correlations.} 
To explore whether the bispectra arising from different spinning particles can be distinguished from each other, as well as from the other standard non-Gaussian shapes, we will now study their shape correlations.
Given two shape functions $S_a$ and $S_b$, we define their shape correlator as~\cite{Babich:2004gb}  
\be
 {\cal C}( S_a,S_b) \equiv \frac{\langle S_a, S_b\rangle}{\sqrt{\langle S_a, S_a\rangle\langle S_b, S_b\rangle}}\, ,
\label{eq:CorrPrim}
\ee
where the inner product between two shape functions is defined by
\be
\langle S_a, S_b\rangle \equiv \int_0^1 d x\int_{1-x}^1 d y\,  S_a(1,x,y)S_b(1,x,y)\, .
\ee
This correlator provides a quantitative measure of how (dis)similar two given shapes are, and thus whether they can be distinguished in actual observations.

\begin{figure}[t!]
	\centering
	\hspace{-.3in}\includegraphics[scale=1.1]{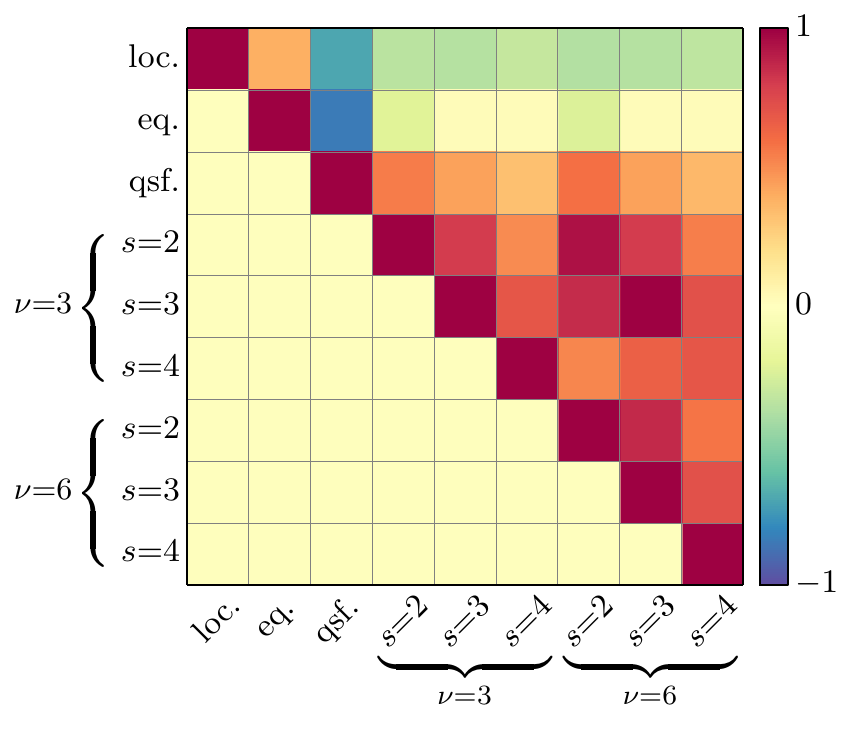}	
	\caption{Shape correlation between the local \eqref{Blocal}, equilateral \eqref{Beq}, quasi-single-field \eqref{BQSF} non-Gaussianity, and the bispectrum template \eqref{eq:Bfull} for $s=\{2,3,4\}$ and $\nu=\{3,6\}$.}	\label{fig:corr}
\end{figure}

For the rest of the text, we will focus on the cases $s=\{2,3,4\}$, and consider two particular mass values corresponding to $\nu=\{3,6\}$ as representatives for the low- and high-mass regimes. Figure~\ref{fig:corr} shows shape correlations between the usual local, equilateral, quasi-single-field shapes of non-Gaussianity, and the bispectrum template with different masses and spins. 
From this figure, we can see that the template is fairly uncorrelated of the three usual shapes of non-Gaussianity, with only moderate correlation ($|{\cal C}|\sim 0.5$) with local-type for the high-$\nu$ case.
Moreover, among the spinning bispectra, we see that adjacent spins retain some correlation, for fixed $\nu$, which is diminished as the two spins are farther apart. Given that this correlation is never too close to unity (for instance, ${\cal C} \sim 0.8$ between spins 2 and 3 for $\nu=3$), we infer that signals from different spins can in principle be distinguished, provided a significant-enough detection.
Finally, we see that same-spin signals with different values of $\nu$ can be fairly correlated, especially in the $s=2$ and 3 cases. 
This is to be expected, given that
very-massive particles (with $\nu\gg1$) produce a signal that asymptotes to the analytic part of Eq.~\eqref{eq:Bfull}, making it impossible to distinguish different masses. We will quantify more precisely how well we can distinguish the particle masses in our forecasts, and address this issue.
We invite the reader to visit Appendix~\ref{sec:AppShapes}, where we expand this analysis to higher-spin cases, albeit only in the $\nu\gg1$ limit.

\section{The Observed Galaxy Bispectrum}
\label{sec:Galaxies}

The material in this section serves largely as a review of existing works in the literature. For the reader familiar with the topic, we first present the expression of the bispectrum that we use in our forecast  (neglecting the Alcock-Paczynski (AP) effect here). We discuss in more details the ingredients of the modeling (biasing, redshift-space distortions, and the AP effect) in the following subsections.  

We model the galaxy bispectrum in redshift space, at leading order in perturbation theory, accounting for the AP effect.  The total bispectrum can be schematically written as 
\begin{align}\label{eq:sum}
B_g(\k_1,\k_2,\k_3,z) \equiv B_g^{\rm grav}(\k_1,\k_2,\k_3,z)  + B_g^{\rm PNG}(\k_1,\k_2,\k_3,z)\, ,
\end{align}
where the first term, $B_g^{\rm grav}$, is due to nonlinear gravitational evolution and the second term, $B_g^{\rm PNG}$, is due to the non-Gaussian initial conditions. Although the above bispectrum is written in terms of three vectors $\k_1$, $\k_2$, $\k_3$, with 9 degrees of freedom, the momentum conservation, $\k_1 + \k_2 +\k_3 =0$, reduces the number of independent variables to 6. 
In the absence of any anisotropy, among the 6 variables, 3 are redundant and the bispectrum can be expressed in terms of three numbers $k_1$, $k_2$, $k_3$, i.e., the sides of the triangle. 
However, Fourier modes are distorted depending on their orientation with respect to the line of sight, due to RSD and the AP effect. 
Therefore, to characterize the galaxy bispectrum, in addition to the shape of the triangles, we need two more angles to specify the position of the triangle with respect to the line of sight.
We can choose these two angles to be the angle $\theta_1$ between the vector $\k_1$ and the line-of-sight direction, $\hat\n$, defined through $\cos \theta_1 = \hat \k_1\cdot \hat\n$, and the azimuthal angle $\phi$, between the vectors $\k_1$ and $\k_2$ ~\cite{Scoccimarro:2015bla,Gagrani:2016rfy}. Following Ref.~\cite{Yamamoto:2016anp}, we define the vectors $\k_1$, $\k_2$, and the line-of-sight direction $\hat\n$, in terms of these $5$ variables as  
\begin{align}
\k_1 &= (0,0,k_1)\, , \quad \k_2 = (0, k_2 \sin \theta_{12}, k_2 \cos \theta_{12})\, , \quad \hat\n = (\sin \theta_1 \cos \phi, \sin \theta_1 \sin \phi, \cos \theta_1)\, . 
\end{align}
Therefore, the angles between $\k_i$ and the line of sight can be written in terms of $\mu_1$ and $\phi$ as
\be\label{eq:mus}
\mu_1 = \cos \theta_1 = \hat \k_1\cdot \hat\n,  \quad  \mu_2 = \mu_1 \cos \theta_{12} + \sqrt{1-\mu_1^2} \ {\rm sin}\theta_{12} \sin \phi, \quad \mu_3 = -\frac{k_1}{k_3}\mu_1-\frac{k_2}{k_3}\mu_2\, ,
\ee
with $\mu_i \equiv \hat\k_i\cdot\hat\n$ and $\cos\theta_{12}\equiv \hat\k_1\cdot\hat\k_2$.

Neglecting the AP effect, at tree-level in perturbation theory, the gravity-induced bispectrum in redshift space is given by \cite{Verde:1998zr, Scoccimarro:1999ed, Scoccimarro:2000sn} 
\begin{align}\label{eq:Bgrav}
	\hskip -6.5pt B_g^{\rm grav}\!(\k_1,\k_2,\k_3,z) = D_{\rm FoG}^B(\k_1,\k_2,\k_3)[2Z_1(\k_1)Z_1(\k_2)  Z_2(\k_1,\k_2)P_0(k_1,z)P_0(k_2,z) + \text{perms}]\, .
\end{align}
The kernels $Z_i$ are the redshift-space perturbation-theory kernels given in Eq.~\eqref{eq:Zfunction} and $D_{\rm FoG}^B$  is the Finger-of-God suppression factor for the bispectrum given in Eq.~\eqref{eq:FOG}. $P_0$ is the matter power spectrum linearly extrapolated to redshift $z$. For primordial non-Gaussianity with the bispectrum given by the template $B_\zeta$, the tree-level contribution to the galaxy bispectrum is 
\begin{align}\label{eq:Bfullgal}
	B_g^{\rm PNG}(\k_1,\k_2,\k_3,z) = D_{\rm FoG}^B(\k_1,\k_2,\k_3) \prod_{i=1}^3\left[ Z_1(\k_i){\mathcal M}(k_i,z) \right] B_\zeta(k_1,k_2,k_3)\, ,
\end{align}
where ${\cal M}(k,z)$ is the transfer function that relates the primordial fluctuations $\zeta$ to the linearly extrapolated matter overdensity $\delta_0$ during the matter-domination era, 
\begin{align}
	\delta_0(\k,z) = {\mathcal M}(k,z)\zeta(\k)\, , \quad
{\mathcal M}(k,z) = -\frac{2}{5} \frac{k^2 T(k)D(z)}{\Omega_m H_0^2}\, ,
\end{align}
with $T(k\rightarrow 0) =1$. As mentioned before, in addition to the template for the bispectrum due to spinning particles, given in Eq.~\eqref{eq:Bfull}, we also consider the local \eqref{Blocal}, equilateral \eqref{Beq}, and quasi-single-field \eqref{BQSF} templates. The galaxy power spectrum, which is needed in calculating the Fisher matrix, at this order is 
\begin{align}\label{eq:PS}
P_g(k,\mu,z) = D_{\rm FoG}^P(\k) Z^2_1(\k)P_0(k,z)\, ,
\end{align}
where $D_{\rm FoG}^P$  is the power spectrum Finger-of-God suppression factor, given in Eq.~\eqref{eq:FOG}. The expression in Eqs.~(\ref{eq:Bgrav}), (\ref{eq:Bfullgal}), and (\ref{eq:PS}) are further modified due to the Alcock-Paczynski effect. The final expressions accounting for this effect is presented in Section~\ref{sec:AP}. 

Before describing the details of the model, let us make a few remarks regarding the importance of loop contributions due to both PNG and gravitational evolution. 
For massive spinning particles, we expect the tree-level contribution from primordial bispectrum to carry most of the information, since it preserves the angular dependence induced by the spin of particles. Therefore, we safely neglect loop corrections on the PNG side. 
Nonetheless, it has been shown in several works that the tree-level treatment of the gravitational bispectrum ceases to be a good approximation well below non-linear scales (see \cite{McEwen:2016fjn, Schmittfull:2016jsw, Schmittfull:2016yqx, Simonovic:2017mhp} for recent works on efficient computation of loop corrections). Moreover, small-scale non-linearities, outside the validity regime of perturbation theory, can backreact on larger scales \cite{Baumann:2010tm,Carrasco:2012cv, Pajer:2013jj,Manzotti:2014loa,Porto:2013qua,Vlah:2015sea}. Hence, for a realistic forecast, one needs to account for the loop corrections to the bispectrum in addition to non-perturbative corrections due to small scales. Our goal here is to provide a first forecast on detectability of the signal from massive particles with nonzero spin during inflation, and we leave an improvement in modeling of the galaxy bispectrum to future work. We refer the reader to Refs.~\cite{Baldauf:2016sjb,Welling:2016dng}, where the impact of the uncertainties in theoretical modeling of the LSS bispectrum on constraints on PNG is studied.

\subsection{Galaxy Bias}
We assume a simple bias model in Eulerian space, where the galaxy overdensity $\delta_g$ at a point $\mathbf x$ can be expanded in terms of the matter overdensity $\delta_m$ and the traceless part of the tidal tensor at the location $\mathbf x$ \cite{McDonald:2006mx, McDonald:2009dh, Assassi:2014fva} (see \cite{Desjacques:2016bnm} for an extensive review). Up to quadratic order, we can write
\begin{align}\label{eq:bias_expansion}
	\delta_g = b_1 \delta_m + \frac{1}{2} b_2 \delta_m^2 +  \frac{1}{2}  b_{K^2}\left[K_{ij}\right]^2, 
\end{align}
where $K_{ij}$ is the tidal tensor, defined as 
\begin{align}
	K_{ij}(\x) \equiv \left( \frac{\partial_i\partial_j}{\partial^2} - \frac{1}{3} \delta_{ij} \right) \delta_m(\x)\, .
\end{align}
We further assume that, in the presence of primordial non-Gaussianity, the bias expansion is not altered. This is a valid assumption for primordial non-Gaussianity due to spinning particles, as the corrections to the linear bias are suppressed, and do not show a distinct scale-dependence on large scales \cite{MoradinezhadDizgah:2017szk}.  
 
\subsection{Redshift-Space Distortions}\label{sec:RSD}

Galaxies are observed in redshift space, as opposed to real space, which
causes their peculiar velocities to  modify their inferred redshift and, hence, their observed distribution. Qualitatively, there are two kinds of contributions to redshift-space distortions. On large scales, the coherent infall velocities of galaxies into high-density regions enhance the clustering amplitudes, while on small scales, within virialized objects, the randomness of galaxy velocities causes de-phasing and leads to a suppression of clustering. The former can be calculated perturbatively, while the latter is a non-perturbative effect (see \cite{Scoccimarro:1999ed,Smith:2007sb,Gil-Marin:2014pva,Tellarini:2016sgp,Yamamoto:2016anp} for a non-exhaustive list of previous works on modeling the redshift-space galaxy bispectrum based on fitting formulae or analytical treatments using perturbation theory or halo model).

The mapping between the redshift- and real-space positions, $\s$ and $\x$, is given by 
\be
\s = \x + \frac{(1+z) v_n(\x)}{H(z)}\, \hat \n\, ,
\ee
where $v_n(\x)$ is the component of the peculiar velocity of the galaxy along the line of sight and $H(z)$ is the Hubble parameter at redshift $z$. The galaxy density contrast in redshift space, $\delta_g^s(\s)$, can then be related to that in real space, $\delta_g(\x)$, using mass conservation as
\be
\left[1+\delta_g^s(\s)\right]d^3s = \left[1+\delta_g(\x)\right]d^3 x\, ,
\ee 
which leads to 
\begin{align}
\delta_g^s(\s) &= \left| \frac{\partial \s}{\partial \r}\right|^{-1} \left[1+\delta_g(\x)\right] - 1\, , \nonumber \\
&=  \left[ 1- f\, \hat\n\cdot{\bf \nabla} u_n(\x) \right]^{-1} \left[1+\delta_g(\x)\right] - 1\, ,
\end{align}
where $f\equiv d\ln D(z)/ d\ln a$ is the logarithmic derivative of the linear growth factor $D(z)$, and we introduced the normalized peculiar velocity along the line of sight as $u_n  = - v_n/(aHf)$. In writing the second line, the distant-observer approximation is used in expressing the Jacobian. The Fourier transform of the density contrast in redshift space is then given by 
\begin{align}\label{eq:RS_delta}
\delta_g^s(\k) &\equiv \int d^3s \,  e^{-i\k\cdot\s} \,  \delta_g^s(\s) \nonumber \\
&= \int d^3 x\, e^{i\k\cdot(\x - f u_n \hat\n )} \left[\delta(\x) + f\, \hat\n\cdot {\bf \nabla} u_n(\x) \right].
\end{align}
In linear perturbation theory, this reduces to the Kaiser effect \cite{Kaiser:1987qv}, i.e., an enhancement of the linear density contrast in redshift space with respect to that in real space
\be
\delta_g^s(\k) = \delta_g(\k) (1+f\mu^2)\, ,
\ee
where $\mu$ is the angle between the wavevector $\k$ and the line-of-sight direction $\hat\n$.  Note that Eq.~\eqref{eq:RS_delta} is the full non-linear expression for the density field in redshift space. In what proceeds, we assume factorizable forms for distortions on both large and small scales.  We use the perturbative evaluation of the above expression to account for the large-scale effects and account for the non-linear effect due to random pairwise velocities as a phenomenological damping factor, which we discuss below. 

 Assuming the bias expansion in Eq.~\eqref{eq:bias_expansion}, perturbative solutions for the density contrast in redshift-space, can be found order by order \cite{Verde:1998zr,Scoccimarro:1999ed,Bernardeau:2001qr}, 
\be 
\delta_g^s(\k) = \sum_{n=1} \int \frac{d^3q_1}{(2\pi)^3} \cdots \int \frac{d^3q_n}{(2\pi)^3}\, \delta_D(\k-\q_{1\cdots n}) Z_n(\q_1,\cdots,\q_n) \delta_0(\q_1) \cdots \delta_0(\q_n)\, ,
\ee
where $\q_{1\cdots n} = \q_1+\q_2+\cdots +\q_n$, $\delta_0$ is the linear matter density contrast, and $Z_n$ are the redshift-space perturbation-theory kernels.  The two lowest-order kernels, which appear in the expression of the tree-level bispectrum can be expressed as 
\begin{align}\label{eq:Zfunction}
	Z_1(\k_1) &= b_1+f \mu_1^2 \, ,\nonumber \\
	Z_2(\k_1,\k_2) &= \frac{b_2}{2} + b_1 F_2(\k_1,\k_2) + f\mu_3^2\hs G_2(\k_1,\k_2) \nonumber \\
	&- \frac{f \mu_3 k_3}{2}\left[ \frac{\mu_1}{k_1} (b_1+f\mu_2^2) +\frac{\mu_2}{k_2}(b_1+f\mu_1^2)\right] + \frac{b_{K^2}}{2}K_2(\k_1,\k_2) \, ,
\end{align}
in terms of the nonlinear kernels $F_2$ and $G_2$ of matter density and velocity contrasts~\cite{Fry:1983cj,Goroff:1986ep,Jain:1993jh,Bernardeau:2001qr}
\begin{align}
	F_2(\k_1,\k_2) &\equiv \frac{5}{7} + \frac{\k_1.\k_2}{2k_1k_2}\left(\frac{k_1}{k_2} + \frac{k_2}{k_1}\right) + \frac{2}{7}  \left(\frac{\k_1.\k_2}	{k_1k_2}\right)^2\, , \nonumber \\
	G_2(\k_1,\k_2) &\equiv \frac{3}{7} + \frac{\k_1.\k_2}{2k_1k_2}\left(\frac{k_1}{k_2} + \frac{k_2}{k_1}\right) + \frac{4}{7} \left(\frac{\k_1.\k_2}{k_1k_2}\right)^2\, ,
\end{align}
and $K_2$, which is the square of the tidal field in Fourier space
\begin{align}
	K_2(\bk_1,\bk_2) \equiv \left(\frac{\bk_1.\bk_2}{k_1 k_2}\right)^2 - \frac{1}{3}\, .
\end{align}

To model the FoG effect, we assume a Gaussian form both for the power spectrum and bispectrum (e.g., \cite{Peacock:1993xg,Percival:2008sh,Song:2008qt, Taruya:2010mx, Hashimoto:2017klo}),
\begin{align}\label{eq:FOG}
D_{\rm FoG}^P(\k) &= {\rm exp} \left[-\frac{k^2\mu^2 \sigma_v^2(z)}{H^2(z)}\right],\quad D_{\rm FoG}^B(\k_1,\k_2,\k_3)  = {\rm exp} \left[-\frac{\sum_{i=1}^3 k_i^2\mu_i^2 \sigma_v^2(z)}{H^2(z)}\right],
\end{align} 
where we have accounted for the effective velocity dispersion in each redshift bin due to the redshift uncertainty $\sigma_z$ of the survey \cite{Asorey:2012rd},
\be
\sigma_v^2(z)=(1+z)^2\left[\frac{\sigma_{\rm FoG}^2(z)}{2} + c^2 \sigma_z^2\right]. 
\ee
Following Ref.~\cite{Giannantonio:2011ya}, we take the redshift dependence for the FoG effect to be
\be
\sigma_{\rm FoG}(z) = \sigma_{{\rm FoG},0}\sqrt{1+z}\, .
\ee
Observations of red and blue galaxies indicate that on average, early-type, red galaxies, which reside in more massive virialized regions, have larger velocity dispersion compared to star-forming blue galaxies, which typically reside in lower-mass halos \cite{Lahav:2002tj,Zehavi:2001nr, Coil:2007jp, Cabre:2008ta,Guo:2012nk}. Therefore, one expects larger values of $\sigma_{\rm FoG}$ for red galaxies. Given the large uncertainties in the measurement of pairwise velocities, in our forecast, we consider $\sigma_{{\rm FoG},0}$ as a free parameter and marginalize over it. 

\subsection{Alcock-Paczynski Effect}\label{sec:AP}

The Alcock-Paczynski effect arises from assuming a fiducial cosmology to infer the position of a galaxy from its observed redshift and angle. If the `true' background cosmology differs from the fiducial cosmology, the inferred comoving radial and transverse distances are distorted with respect to the true ones. 
We can then relate the observed galaxy spectra of the reference and true cosmologies as \cite{Seo:2003pu, Song:2015gca}
\begin{align}
B_g^{\rm obs}(\tilde k_1, \tilde k_2, \tilde k_3,\tilde \mu_1,\tilde \mu_2) &= \left[\frac{H_{\rm true}(z)}{H_{\rm ref}(z)}\right]^2\left[\frac{D_{A,{\rm ref}}(z)}{D_{A,{\rm true}}(z)}\right]^4 B_g(k_1,k_2,k_3,\mu_1,\mu_2)\, , \\[4pt]
P_g^{\rm obs}(\tilde k, \tilde \mu) &=  \frac{H_{\rm true}(z)}{H_{\rm ref}(z)} \left[\frac{D_{A,{\rm ref}}(z)}{D_{A,{\rm true}}(z)}\right]^2 P_g(k,\mu)\, ,
\end{align}
where the tilde coordinates are those in the reference cosmology.
These can be related to the `true' ones through
\begin{align}
k_i &= \tilde k_i \left[ (1- {\tilde \mu}_i^2) \frac{D_{A,{\rm ref}}^2(z)}{D_{A,{\rm true}}^2(z)} + {\tilde \mu}_i^2 \frac{H_{\rm true}^2(z)}{H_{\rm ref}^2(z)}\right]^{1/2} , \quad \mu_i =  {\tilde \mu}_i \frac{\tilde k_i}{k_i} \frac{H_{\rm true}(z)}{H_{\rm ref}(z)}\, ,
\end{align}
where $D_A(z)$ is the angular-diameter distance. 
The distance ratios account for the difference in volume between the two cosmologies.

\section{Fisher Forecasts}
\label{sec:Fisher}
In this section, we present forecasts for the potential of upcoming galaxy surveys to constrain primordial non-Gaussianity due to the presence of higher-spin particles based on Fisher analysis of the galaxy bispectrum. We will consider two spectroscopic surveys: DESI and EUCLID. {\bf We also discuss the impact of theoretical error and non-Gaussian corrections to the bispectrum covariance on the forecasted constraints on this type of non-Gaussianity.}

In general, the Fisher matrix is defined as~\cite{Tegmark:1997rp}
\be
F_{\alpha \beta} = - \left< \frac{\partial^2\hs {\rm ln}\hs {\mathcal L}({\bf x} | \boldsymbol{\lambda})}{\partial \lambda_\alpha \partial \lambda_\beta}\right>,
\ee
where  ${\mathcal L}$ is the likelihood of the data ${\bf x}$ given the parameters $\boldsymbol {\lambda}$.  
The Fisher matrix provides an easy and accurate approximation to the errors that can be achieved by an experiment, assuming that the likelihood $\cal L$ is Gaussian.
In that case, the forecasted marginalized uncertainty in the $\alpha$-th parameter is found through $\sigma^2 (\lambda_{\alpha}) = \left({\bf F}^{-1}\right)_{\alpha \alpha}$.
For reference, this uncertainty is always larger than the unmarginalized uncertainty $1/\sqrt{F_{\alpha \alpha}}$, as
a consequence of the Cram\'er-Rao inequality~\cite{Creminelli:2006gc,Kamionkowski2011}.

The Fisher matrix of the galaxy bispectrum at a given redshift bin with mean $z_i$ is given by
\begin{align}
	F_{\alpha \beta}(z_i) = \frac{V_i}{(2\pi)^5} \int_{{\mathcal V}_B}  d V_k\, k_1 k_2 k_3\int_{-1}^1  d \cos \theta_1 \int_0^{2\pi}\d\phi\, \frac{(\partial B_g^{\rm obs}/\partial\lambda_\alpha)(\partial B_g^{\rm obs}/\partial\lambda_\beta)}{{\rm Var}\hs B_g}\, ,
\label{eq:FisherG}
\end{align}
where we have defined $ d V_k \equiv  d k_1  d k_2  d k_3$ and  ${\mathcal V}_B$ is the tetrahedral domain allowed by the triangle condition for the wavenumbers $k_{\rm min} < k_i< k_{\rm max}$.  The volume of the redshift bin in the range,  $z_{\rm min} < z_i < z_{\rm max}$, for a  survey covering a fraction of sky $f_{\rm sky}$ is given by
\be
V_i=\frac{4\pi f_{\rm sky}}{3} \big[d_c^3(z_{\rm max}) -d_c^3(z_{\rm min}) \big]\, ,
\ee
with $d_c$ the comoving distance to redshift $z$,
\be
d_c(z) = \int_0^z d z\, \frac{c}{H(z)} \, .
\ee
The variance of the galaxy bispectrum  
is given by 
\begin{align}
	{\rm Var}\hs B_g(k_1,k_2,k_3,\theta_1,\phi,z_i)
	 & = s_{123} \prod_{j=1}^3\left[P_g^{\rm obs}(k_j,\mu_j,z_i) + \frac{1}{\bar n_i}\right],
\end{align}
where $s_{123}=6, 2, 1$ for equilateral, isosceles, and scalene triangles, respectively, $\mu_1,\mu_2$ and $\mu_3$ are given by Eq. \eqref{eq:mus}, and $\bar n_i$ is the shot noise in the $i$-th redshift bin. The total Fisher matrix will be obtained by summing the Fisher matrices over all the redshift bins,
\be
F_{\alpha \beta} = \sum_i F_{\alpha \beta}(z_i)\, .
\ee
For each redshift bin,  we set $k_{\rm min} = 2\pi/V_i^{1/3}$,
and we set the value of $k_{\rm max}$ such that the variance of the linear density field at that redshift is
\be
\sigma^2(z) = \int_{k_{\rm min}(z)}^{k_{\rm max}(z)} \frac{d^3 k}{(2\pi)^3}\, P_0(k,z) = \sigma^2(z=0)\, ,
\label{eq:kmax}
\ee 
where $ \sigma^2(z=0)$ is chosen to satisfy $k_{\rm max}(z=0) \simeq 0.15$ $h$ Mpc$^{-1}$.
In order to test the dependence of our results on $k_{\rm max}$, as well as to present a conservative estimate well within the range of validity of our tree-level bispectrum model, we will also quote results for $k_{\rm max}=0.075\hs h$ Mpc$^{-1}$ at $z=0$, which are obtained through the same procedure, albeit changing the right-hand side of Eq.~\eqref{eq:kmax}.

In our analysis, we obtain constraints on the amplitude of primordial non-Gaussianity $f_{\rm NL}$, due to particles with spins $s=2,3,4$ and their corresponding masses expressed in terms of the parameter $\nu$. We also vary five $\Lambda$CDM cosmological parameters: the amplitude $\ln(10^{10}A_s)$ and the spectral index $n_s$ of primordial fluctuations, the Hubble parameter $h$, the energy density $\Omega_{\rm cdm}$ of cold dark matter, and  that of baryons $
\Omega_b$. We vary three biases, $b_1, b_2$ and $b_{K^2}$, and a single dispersion velocity $\sigma_{\rm FOG}$ for the FOG effect. Therefore, given a particle spin, our parameter arrays are given by: ${\boldsymbol\lambda} = \left[{\rm ln} (10^{10}A_s), n_s, h,\Omega_{\rm cdm},\Omega_b, f_{\rm NL}, \nu, \sigma_{{\rm FOG},0}, b_1, b_2, b_{K^2} \right]$.

For the fiducial values of cosmological parameters, we choose $\ln(10^{10}A_s) = 3.067, n_s = 0.967, h = 0.677$, $\Omega_{\rm cdm} =0.258$, $\Omega_b = 0.048$, and we set  the pivot scale to be $k_p=0.05 \ {\rm Mpc}^{-1}$, consistent with Planck 2015 data \cite{Ade:2015xua}. We use the public CLASS code \cite{Lesgourgues:2011re,Blas:2011rf} to compute the linear matter power spectrum. We set the fiducial values for the parameters characterizing PNG to be $f_{\rm NL}= 1$, and $\nu=3$. When considering the quasi-single-field model, we take $\tilde \nu = 1$, since the template of QSF is valid for $0<\tilde\nu<3/2$. The projected constraints on $f_{\rm NL}$ are only marginally affected by its fiducial value, whereas constraints on $\nu$ are significantly altered. We expect the constraints on $\nu$ to scale as $\sigma(\nu) \propto 1/f_{\rm NL}$, which we have checked with an additional run with a fiducial value of $f_{\rm NL}=10$. To study the dependence of our results on the fiducial value of $\nu$, we also present the constraints for a fiducial value of $\nu=6$.  We set the fiducial value of the velocity dispersion for both surveys to be $\sigma_{{\rm FOG},0} = 250 \ {\rm km }$ s$^{-1}$ \cite{Giannantonio:2011ya, Lahav:2002tj,Zehavi:2001nr,Cabre:2008ta,Guo:2012nk}.

We model the redshift evolution of the linear bias as $b_1(z) = \bar b_1 p(z)$, where $\bar b_1$ is a free amplitude that we vary and $p(z)$ captures the redshift  dependence. We set the fiducial value of $\bar b_1 = 1.46$ such that at $z=0$ the value of the linear bias is consistent with the results in Ref.~\cite{Lazeyras:2015lgp} for halos of mass $M = 3\times 10^{13} h^{-1} M_\odot$. For EUCLID, we consider $p(z) = \sqrt{1+z}$  \cite{Rassat:2008ja}, while for DESI we take $p(z) = 0.84/D(z)$ where $D(z)$ is the growth factor normalized to unity at $z=0$. For the fiducial values of quadratic biases we assume the scaling relations of $b_2 = \bar b_2(0.412 -  2.143 b_1 + 0.929 b_1^2 + 0.008 b_1^3)$ and $b_{K^2} = \bar b_{K^2}(0.64 - 0.3 b_1 + 0.05 b_1^2 -0.06 b_1^3)$, which are fits to N-body simulations provided in Refs.~\cite{Lazeyras:2015lgp,Modi:2016dah}. Based on these results, we assume that the above relation between $b_2$ and $b_{K^2}$ with $b_1$, is preserved in the redshift range we consider and use it to set the fiducial values of the the biases in each redshift bin. We vary two parameters for the overall amplitudes $\bar b_2$ and $\bar b_{K^2}$.

\subsection{Survey Specifications}

\begin{figure}[t!]\centering
\includegraphics[scale=.7]{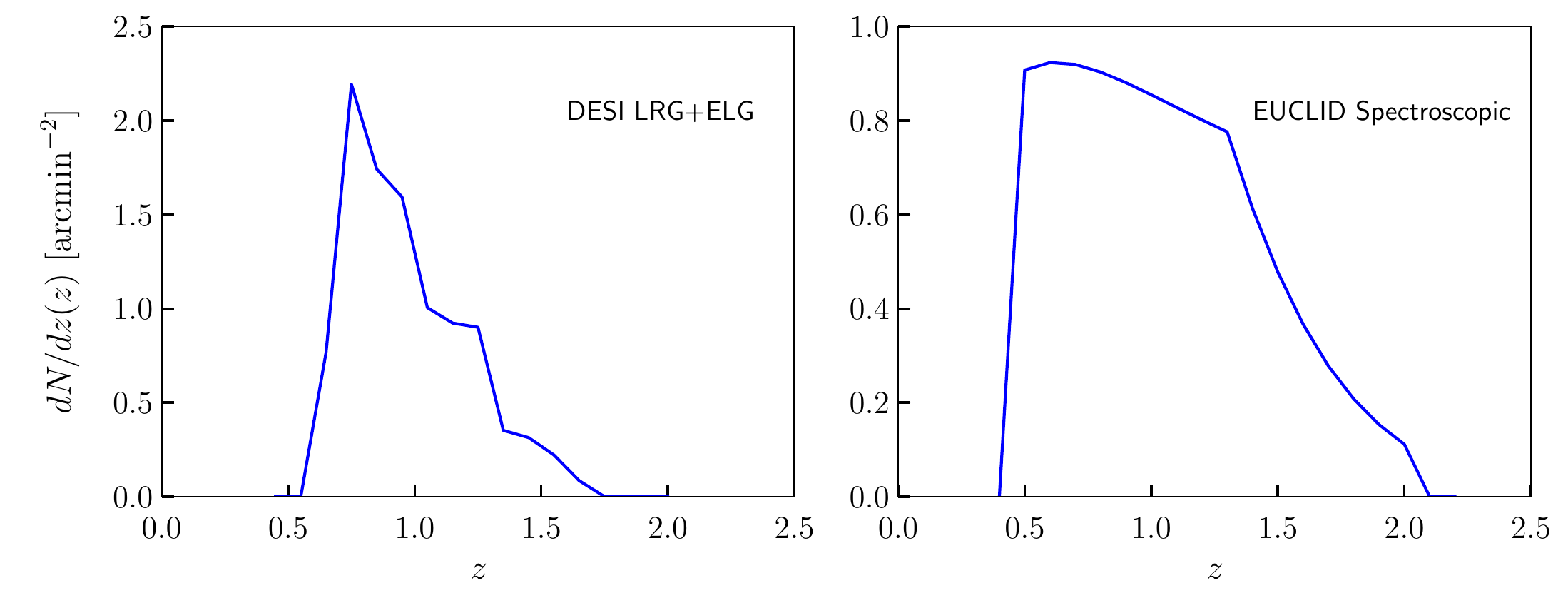}
\caption{Normalized redshift distribution of galaxies for DESI \cite{Aghamousa:2016zmz} (\emph{left}) and EUCLID spectroscopic survey  \cite{Geach:2009tm} (\emph{right}). For DESI, the distribution corresponds to the sum of ELG and LRG galaxies. For EUCLID, the distribution is obtained from empirical data of the luminosity function of H$\alpha$ emitters out to $z=2$. We take the limiting flux to be $4\times 10^{-16} {\rm erg \ s}^ {-1} {\rm cm}^{-2}$ and an efficiency of $35\%$.}
\label{fig:dNdz}
\end{figure}

In our analysis, we assume top-hat redshift bins and consider them to be uncorrelated with each other. Therefore the shot-noise in redshift bin $i$ is given by
\be
\bar n_i = \frac{4\pi f_{\rm sky}}{V_i} \int_{z_{\rm min}}^{z_{\rm max}}  d z \, \frac{ d N}{ d z}(z)\, ,
\ee
where $dN/dz$ is the surface number density of a galaxies in the survey, which is shown in Fig.~\ref{fig:dNdz} for the two surveys we consider.  Let us briefly comment on the assumption of uncorrelated bins. In principle, there are two sources for correlation between the bins, first is due to gravitational clustering and the second, due the error in the redshift estimates. While the former is an additional source of signal, the latter is a source of noise. Since for spectroscopic surveys, uncertainties in redshift estimates are small, the cross-correlation between bins due to redshift errors can be safely neglected. 

For the forecasts done in this work, we consider the following two surveys:
\begin{itemize}
\item {\bf EUCLID:}  EUCLID \cite{Laureijs:2011gra} is a European Space Agency medium-class mission expected to launch in 2020. It will measure two complementary observables: galaxy clustering and weak gravitational lensing. We will only consider  the galaxy clustering part of the mission, and assume a sky fraction of $f_{\rm sky} = 0.36$, corresponding to a coverage of $15,000 \ {\rm deg}^2$. We take 12 equally populated redshift bins in the range $0.4<z<2.1$, similar to the procedure in Ref.~\cite{Giannantonio:2011ya}. We assume the redshift uncertainty to be $\sigma_z(z) = 0.001(1+z)$. We use the redshift distribution $dN/dz$ given by the tabulated data in Ref.~\cite{Geach:2009tm}, obtained from empirical data of luminosity function of H$\alpha$ emitters (see Ref.~\cite{Pozzetti:2016cch} for an updated empirical model using a larger data combination). We take the limiting flux to be $4\times 10^{-16} {\rm erg \ s}^ {-1} {\rm cm}^{-2}$, and an efficiency of $35\%$.

\item{\bf DESI:} Dark Energy Spectroscopic Instrument \cite{Aghamousa:2016zmz} is a galaxy and quasar redshift survey. It is expected to run over a five-year period from 2018 to 2022. It will obtain optical spectra for three types of objects: Emission Line Galaxies (ELGs), Luminous Red Galaxies (LRGs) and quasars (QSOs). In our analysis, we consider their galaxy sample consisting of ELGs and LRGs. We assume a sky fraction of $f_{\rm sky} = 0.34$, corresponding to a coverage of $14,000 \ {\rm deg}^2$.  We assume the redshift uncertainty to be $\sigma_z(z) = 0.0005(1+z)$. We use the redshift distribution of ELG and LRG samples from Ref.~\cite{Aghamousa:2016zmz} in the redshift range of $ 0.65 \leq z \leq 1.65$.

\end{itemize}

\subsection{Planck Prior}
\label{sec:Planck}

In our final results, we also show the constraints on parameters of interest, imposing Planck priors, which can improve the constraints from clustering statistics of LSS by breaking degeneracies present between different variables.
We follow Ref.~\cite{Munoz:2015fdv}, and use the public Planck data\footnote{\tt{https://pla.esac.esa.int/pla/}} to obtain the observed CMB angular power spectra, $\tilde C_\ell = C_\ell + N_\ell$, where $C_\ell$ is the CMB power spectrum, and $N_\ell$ is the instrumental noise. We will limit ourselves to temperature data, where (ignoring correlations between modes) we can find the Planck Fisher matrix as~\cite{Zaldarriaga:1996xe,Kamionkowski:1996ks}
\be
F_{\alpha \beta}^{\rm Planck} = \dfrac{1}{f_{\rm sky}^{P} } \sum_{\ell=2}^{2500} \dfrac{2}{2\ell+1}  \frac{(\partial C_\ell/\partial \lambda_\alpha) (\partial C_\ell/\partial \lambda_\beta)}{C_\ell^2}\, .
\ee

\begin{table*}[!t]
\hskip 40pt
	\begin{tabular}{| p{22mm} || >{\centering\arraybackslash}  m{22mm}  | >{\centering\arraybackslash} m{22mm} || >{\centering\arraybackslash}  m{22mm}  | >{\centering\arraybackslash} m{22mm} |}
		\hline
		EUCLID &  \multicolumn{2}{c||}{$k_{\rm max}(z=0)=0.15\, h$ Mpc$^{-1}$} & \multicolumn{2}{c|}{$k_{\rm max}(z=0)=0.075\, h$ Mpc$^{-1}$}\\
		\hline \hline
		& $\sigma(f_{\rm NL})$ & $\sigma(\tilde\nu)$ & $\sigma(f_{\rm NL})$ & $\sigma(\tilde\nu)$
		\\
		\hline
		loc.	& 0.38	& $-$	& 0.74	& $-$
		\\
		eq.	& 2.3	&$-$	&  4.1	& $-$
		\\
		qsf ($\tilde\nu_{\rm fid}=1$)	&  2.2	& 1.3	& 4.0	& 2.5
		\\
		\hline \hline
		$\nu_{\rm fid}=3$ & $\sigma(f_{\rm NL})$ & $\sigma(\nu)$ & $\sigma(f_{\rm NL})$ & $\sigma(\nu)$	
		\\ \hline		
		$s=2$	& 0.66	& 0.35	& 1.2	& 0.62
		\\
		$s=3$	& 1.5	& 3.3	& 2.4	& 5.7
		\\
		$s=4$	& 0.68	& 0.098	& 1.1	& 0.16
		\\	
		\hline 
	\end{tabular}
	\vskip 1pt
	\hskip 40pt
	\begin{tabular}	{| p{22mm} || >{\centering\arraybackslash}  m{22mm}  | >{\centering\arraybackslash} m{22mm} |}
		\hline
		$\nu_{\rm fid}=6$ & $\sigma(f_{\rm NL})$ & $\sigma(\nu)$ 
		\\ \hline
		$s=2$	&  0.71	& 370	
		\\
		$s=3$	& 1.5	& 250	
		\\
		$s=4$	&  0.52	&  0.23
		\\		\hline
	\end{tabular}
	\caption{1-$\sigma$ uncertainties in non-Gaussianity parameters for different models, assuming $\tilde \nu_{\rm fid}=1$,  $\nu_{\rm fid}=3$ and 6, and with
		$f_{\rm NL}=1$ for all cases. Here we have considered the EUCLID survey, and marginalized over all parameters. We show two different cases, labeled by their value of $k_{\rm max}$ at $z=0$.}
\label{tab:fNLE}
\end{table*}
\begin{table*}[!h]
\hskip 40pt	
\begin{tabular}{| p{22mm} || >{\centering\arraybackslash}  m{22mm}  | >{\centering\arraybackslash} m{22mm} || >{\centering\arraybackslash}  m{22mm}  | >{\centering\arraybackslash} m{22mm} |}
		\hline
		DESI &  \multicolumn{2}{c||}{$k_{\rm max}(z=0)=0.15\, h$ Mpc$^{-1}$} & \multicolumn{2}{c|}{$k_{\rm max}(z=0)=0.075\, h$ Mpc$^{-1}$}\\
		\hline \hline
		& $\sigma(f_{\rm NL})$ & $\sigma(\tilde\nu)$ & $\sigma(f_{\rm NL})$ & $\sigma(\tilde\nu)$
		\\
		\hline
		loc.	& 0.53	& $-$	& 1.2	& $-$
		\\
		eq.	& 2.7	&$-$	& 5.3	& $-$
		\\
		qsf ($\tilde\nu_{\rm fid}=1$)	& 2.5	& 1.7	& 5.3	& 3.9
		\\
		\hline	\hline
		$\nu_{\rm fid}=3$	& $\sigma(f_{\rm NL})$ & $\sigma(\nu)$ & $\sigma(f_{\rm NL})$ & $\sigma(\nu)$	
		\\ \hline 
		$s=2$	& 0.80	& 0.48	& 1.6	& 0.89
		\\
		$s=3$	& 1.8	& 4.1	& 3.1	& 8.2
		\\
		$s=4$	& 0.82	& 0.12	& 1.5	& 0.23
		\\				\hline	 
	\end{tabular}

	\vskip 1pt
	\hskip 40pt	
	\begin{tabular}	{| p{22mm} || >{\centering\arraybackslash}  m{22mm}  | >{\centering\arraybackslash} m{22mm} |}
		\hline
		$\nu_{\rm fid}=6$	&  $\sigma(f_{\rm NL})$ &  $\sigma(\nu)$
		\\ \hline	
		$s=2$	&  0.85	&  510	
		\\
		$s=3$	&  1.8	&  320	
		\\
		$s=4$	&   0.64	&  0.30	
		\\		\hline
	\end{tabular}
	\caption{Same as Table~\ref{tab:fNLE}, but for DESI.}
\label{tab:fNLD}
\end{table*}

We take $f_{\rm sky}^{P}=0.6$, and we compute the derivatives $\partial C_\ell/\partial \lambda_\alpha$ by using the publicly available code CAMB~\cite{Lewis:1999bs}.
We add a prior on the optical depth $\tau$ of reionization of $\sigma(\tau) = 0.019$, to represent low-$\ell$ polarization data, and marginalize over $\tau$ before adding the Planck Fisher matrix to the galaxy matrix $F_{\alpha \beta}$.
More explicitly, in the cases that we include a Planck prior our total Fisher matrix will be
\be
F_{\alpha \beta}^{\rm tot} = F_{\alpha \beta} + F_{\alpha \beta}^{\rm Planck}\, ,
\ee
where the Planck matrix only has non-vanishing entries for the five relevant cosmological parameters.

\begin{figure}[!t]
	\centering
	\hspace{-.5in}
	\includegraphics[width=.85\linewidth]{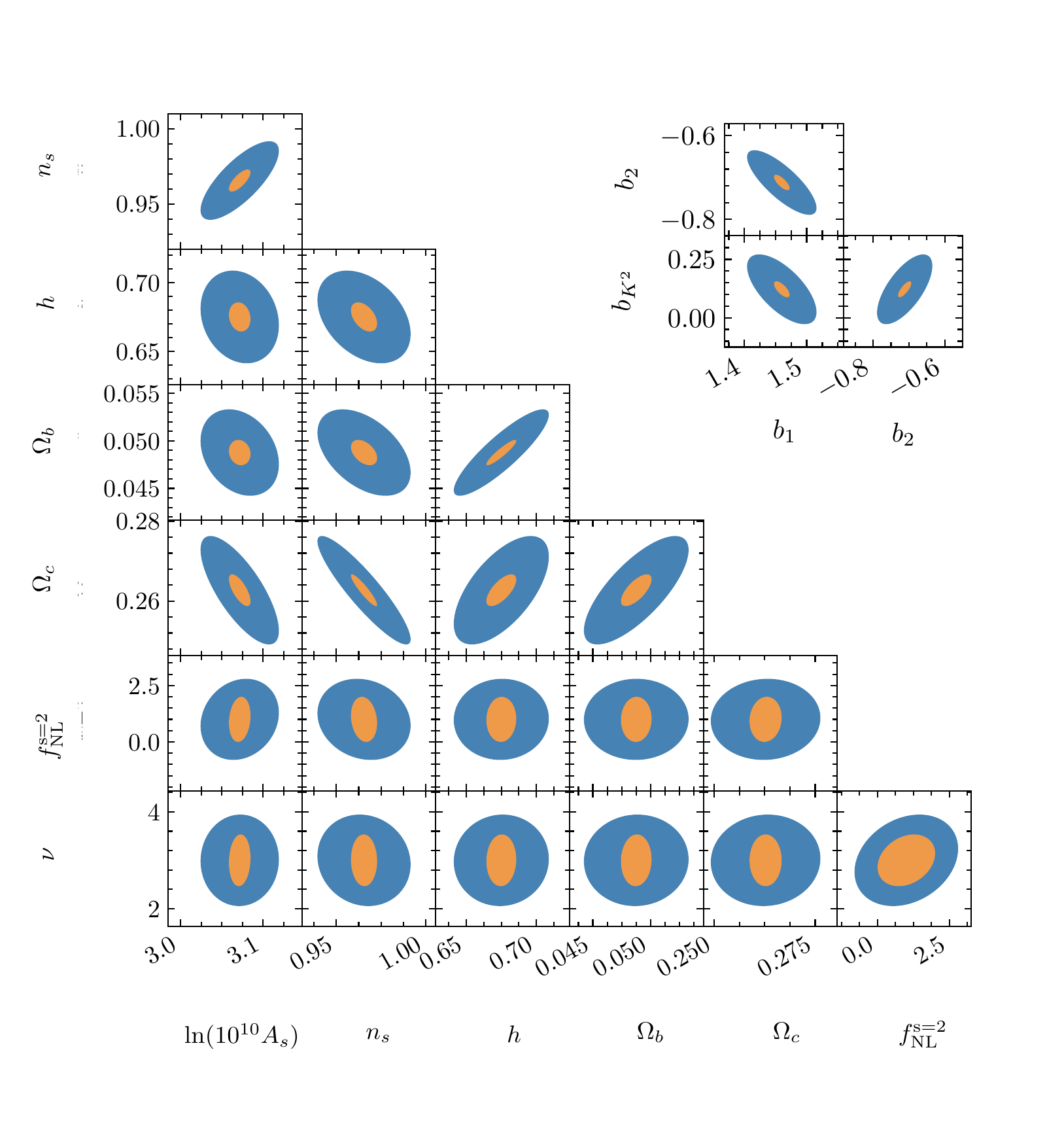}	\vspace{-0.2in}
	\caption{1-$\sigma$ confidence ellipses in $\Lambda$CDM cosmological parameters, as well as the biases and non-Gaussianity parameters, for the EUCLID survey. Our fiducial values are the same than in Table~\ref{tab:fNLE}, which for the non-Gaussianity parameters correspond to $f_{\rm NL}=1$ and $\nu=3$. The orange (inner) and blue (outer) contours correspond to choosing $k_{\rm max}=0.15$ $h$ Mpc$^{-1}$  and $k_{\rm max}=0.075$ $h$ Mpc$^{-1}$ at $z=0$, respectively. }
	\label{fig:Ellipses}
\end{figure}
\begin{figure}{!b}
	\centering
	\hspace{-0.5in}
	\includegraphics[width=0.85\linewidth]{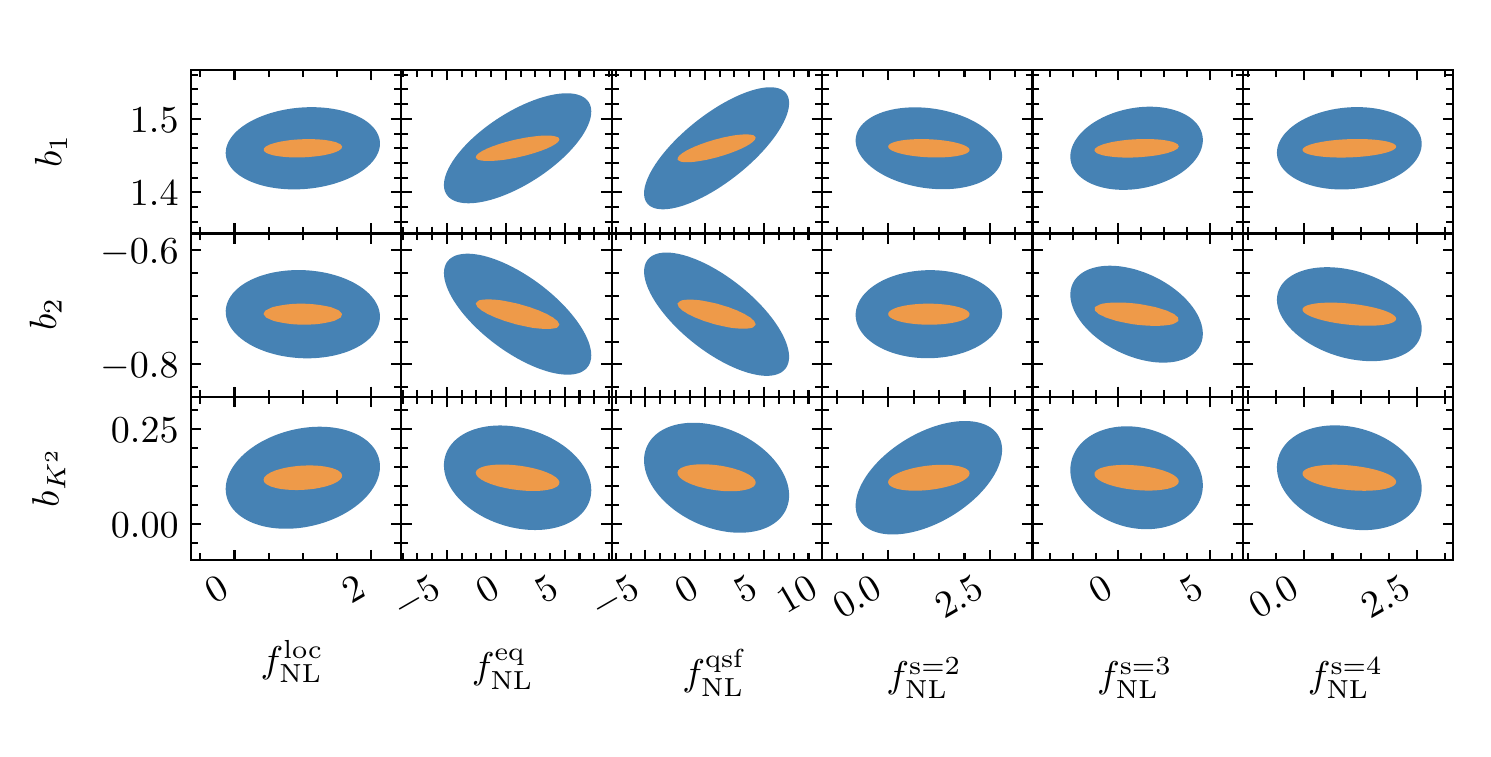}	\vspace{-.2in}
	\caption{$1-\sigma$ confidence ellipses for the biases and the amplitudes of primordial non-Gaussianity for various shapes for the EUCLID survey.
	Our fiducial values are the same as in Table~\ref{tab:fNLE}, which for the non-Gaussianity parameters correspond to $f_{\rm NL}=1$ and $\nu=3$. The orange (inner) and blue (outer) contours correspond to choosing $k_{\rm max}=0.15$ $h$ Mpc$^{-1}$  and $k_{\rm max}=0.075$ $h$ Mpc$^{-1}$ at $z=0$, respectively. }
	\label{fig:Ellipses_bias}
\end{figure}

\subsection{Results}

We present our results of the Fisher forecast in this section. In Tables~\ref{tab:fNLE} and \ref{tab:fNLD}, we show the forecasted errors for parameters of primordial non-Gaussianity for EUCLID and DESI. We show the errors for scenarios with massive particles with spins $s=2,3,4$ as well as  the local, equilateral and quasi-single-field models. To demonstrate how the constraints depend on the smallest scale considered in the forecast, we show the results for two different choices of $k_{\rm max} = 0.15 \ h\, {\rm Mpc}^{-1}$ (which is our default choice for the rest of the results), and a conservative choice of $k_{\rm max} = 0.075 \ h\, {\rm Mpc}^{-1}$. For the largest value of $k_{\rm max}$, we also show the constraints for a different choice of fiducial value of $\nu$. We set a fiducial value of $f_{\rm NL} = 1$ for all models. 

Our results show that both surveys perform comparably, with EUCLID reaching a $\sim 20\%$ better precision on $f_{\rm NL}$ and $\nu$. From these tables, we see that for the local type of non-Gaussianities, these surveys will reach precision below $f_{\rm NL}=1$. Interestingly, the uncertainty in $f_{\rm NL}$ that can be reached for the spinning shapes is also below unity for $s=2$ and $4$, showing the great promise of these surveys to detect spinning fields during inflation. The results are less optimistic for $s=3$, due to the partial cancellation of the odd-spin Legendre polynomial in Eq. \eqref{eq:Bfull}, which leads to a more moderate scaling in the squeezed limit. Additionally, we show how well the masses of the spinning particles can be measured, in terms of $\nu$, for our fiducial case of $f_{\rm NL}=1$.  Interestingly, these errors are small enough that once non-Gaussianities are detected, the particle masses will be determined within a precision of tens of percent. This is not surprising, as different values of $\nu$ give rise to drastically different signatures, which can be distinguished without difficulty. For reference, we also show the constraints that can be achieved using just the galaxy power spectrum in Table~\ref{tab:fNLPowSp} (see Appendix~\ref{sec:PowSp} for details).

For EUCLID, we show in Fig.~\ref{fig:Ellipses} the 1-$\sigma$ ellipses corresponding to correlations of the three bias parameters and $f_{\rm NL}$, for all models of primordial non-Gaussianity that we considered. The 2D-contour plots for the correlations of cosmological parameters with $f_{\rm NL}$ and $\nu$  as well as correlations between the biases are shown in Fig.~\ref{fig:Ellipses_bias}. This figure shows that both $f_{\rm NL}$ and $\nu$ are fairly uncorrelated with the rest of the parameters, although they are somewhat correlated with each other. 
Note that in this figure, the ellipses are obtained for the case of $s=2$ as a representative case. The trends in the correlations are the same for other spins.

As a byproduct of our analysis, we can investigate the potential of the galaxy bispectrum as a probe of the cosmological parameters.
For the case of $s=2$, we show the constraints on cosmological parameters, biases and peculiar velocity of galaxies in Tables \ref{tab:fullE} and \ref{tab:fullD}. The numbers in parentheses are the constraints obtained imposing Planck priors. Our results indicate that for cosmological parameters, linear bias, and peculiar velocity, the constraints from DESI are better by about $10\%$,  while the constraints on higher-order biases, $b_2$ and $b_{\rm K^2}$  from the two surveys are comparable. Once Planck priors are imposed, the two surveys achieve similar constraints. 

	\begin{table*}
		\centering
	\begin{tabular}{| l ||  c   c || c  c |}
		\hline
		EUCLID &  \multicolumn{2}{c||}{$k_{\rm max}(z=0)=0.15$ $h$ Mpc$^{-1}$} & \multicolumn{2}{c|}{$k_{\rm max}(z=0)=0.075$ $h$ Mpc$^{-1}$  } \\             
		\hline \hline
		$\sigma(\ln( 10^{10}\,A_s ))$  & \, $8.6\times 10^{-3}$ \, & ($5.9 \times10^{-3}$) & 	\, $3.2 \times10^{-2}$ \, & ($1.7 \times10^{-2}$) 
		\\
		$\sigma(n_s)$  & $4.8 \times10^{-3}$ & ($2.1 \times10^{-3}$) & $1.7 \times10^{-2}$ & ($3.0 \times10^{-3}$) 
		\\
		$\sigma(h)$  & $7.0\times10^{-3}$ & ($1.1 \times10^{-3}$) &$2.2\times10^{-2}$ & ($2.1\times10^{-3}$)
		\\
		$\sigma(\Omega_b)$ & $8.8 \times10^{-4}$& ($2.3 \times10^{-4}$) &  $3.0 \times10^{-3}$& ($3.7 \times10^{-4}$)
		\\
		$\sigma(\Omega_c)$  & $2.6 \times10^{-3}$ & ($1.1 \times10^{-3}$) & $9.0 \times10^{-3}$ & ($2.5 \times10^{-3}$)
		\\
		$\sigma(\bar b_1)$  & $8.2\times10^{-3}$ & ($7.9\times10^{-3}$) & $3.7\times10^{-2}$ & ($3.0\times10^{-2}$)
		\\
		$\sigma(\bar b_2)$  & $1.2\times10^{-2}$ & ($1.2\times10^{-2}$) & $5.1\times10^{-2}$ & ($4.7\times10^{-2}$) 
		\\	
		$\sigma(\bar b_{K2})$  & $2.3\times10^{-2}$ & ($2.2\times10^{-2}$) & $9.8\times10^{-2}$ & ($9.2\times10^{-2}$)
		\\			
		$\sigma(\sigma_{{\rm FoG},0})$(km s$^{-1}$)  &  1.2 & (1.2)& 8.0 & (7.0)
		\\
		\hline
	\end{tabular}
	\caption{1-$\sigma$ uncertainties in cosmological parameters, as well as for the biases and non-Gaussianity parameters, for EUCLID, for two different cases of $k_{\rm max}$. The numbers in parentheses assume a Planck prior. For the non-Gaussianity parameters we have assumed a spin-2 particle with $f_{\rm NL}=1$ and $\nu=3$.}
\label{tab:fullE}		
\end{table*}
	\begin{table*}
	\centering
	\begin{tabular}{| l || c c || c c |}
		\hline
		DESI &  \multicolumn{2}{c||}{$k_{\rm max}(z=0)=0.15$ $h$ Mpc$^{-1}$} & \multicolumn{2}{c|}{$k_{\rm max}(z=0)=0.075$ $h$ Mpc$^{-1}$  } \\             
\hline \hline
		$\sigma(\ln( 10^{10}\,A_s ))$  & \, $7.5\times10^{-3}$ \, & ($5.3 \times10^{-3}$) & \, $3.2 \times10^{-2}$ \, & ($1.7 \times10^{-2}$) 
		\\
		$\sigma(n_s)$  & $4.5 \times10^{-3}$ & ($2.1 \times10^{-3}$) & $2.0 \times10^{-2}$ & ($3.1 \times10^{-3}$) 
		\\
		$\sigma(h)$  & $6.0\times10^{-3}$ & ($1.1 \times10^{-3}$) &$2.7\times10^{-2}$ & ($2.5\times10^{-3}$)
		\\
		$\sigma(\Omega_b)$ & $8.0 \times10^{-4}$ &($2.3 \times10^{-4}$) &  $3.6 \times10^{-3}$ & ($4.1 \times10^{-4}$)
		\\
		$\sigma(\Omega_c)$  & $2.4 \times10^{-3}$ &($1.1 \times10^{-3}$) & $1.1 \times10^{-2}$ & ($3.2 \times10^{-3}$)
		\\
		$\sigma(\bar b_1)$  & $7.2\times10^{-3}$ & ($7.1\times10^{-3}$) & $3.6\times10^{-2}$ & ($3.0\times10^{-2}$)
		\\
		$\sigma(\bar b_2)$  & $1.2\times10^{-2}$ & ($1.2\times10^{-2}$) & $5.8\times10^{-2}$ & ($5.2\times10^{-2}$) 
		\\	
		$\sigma(\bar b_{K2})$  & $2.3\times10^{-2}$ & ($2.3\times10^{-2}$) & $1.2\times10^{-1}$ & ($1.1\times10^{-1}$)
		\\			
		$\sigma(\sigma_{{\rm FoG},0})$ (km s$^{-1}$)  &  0.68 & (0.66)& 7.9 & (7.1)
		\\		
		\hline
	\end{tabular}
		\caption{Same as Table~\ref{tab:fullE}, but for DESI.}
	\label{tab:fullD}		
\end{table*}

\subsection{Theoretical Error }
In forecasting the constraints from the observed power spectrum and bispectrum, the uncertainty in theoretical modeling should be accounted for as an additional source of error, {\it i.e. theoretical error}~\cite{Baldauf:2016sjb,Welling:2016dng,Byun:2017fkz,Karagiannis:2018jdt}.  In the prescription introduced in Ref.~\cite{Baldauf:2016sjb}, the error covariance matrix can be written as $C_{ij}^e = E_i \rho_{ij} E_j$, where $i, j$ are the indices of different momentum configurations (different triangles for the bispectrum) and $\rho_{i j}$ is the correlation coefficient accounting for correlations between different triangle configurations. $E_i$ is an envelope quantifying the difference between the true spectrum and the model considered, and at each order in perturbation theory is estimated from the next-order term in the loop expansion. In full generality, the theoretical error should be estimated accounting for next-order contributions due to non-linearities in the dark matter distribution, redshift-space distortions, as well as non-linear bias. 
In the presence of primordial non-Gaussianity, next-order loop contributions due to PNG also have to be accounted for. In Ref.~\cite{Baldauf:2016sjb} only next-order loop contributions for matter fluctuations were accounted for. More recently, in Ref.~\cite{Karagiannis:2018jdt} this result was extended to also account for local-in-matter bias terms. A different approach to estimate theoretical error was used in Ref.~\cite{Byun:2017fkz}, where the forecasted constraints on cosmological parameters were compared, when using the predictions of standard perturbation theory at tree and 1-loop levels, as well as that from halo-model versus the measurement of the bispectrum from N-body simulation. 

To quantify the impact of theoretical error on constraints on non-Gaussianity due to massive particles with spin, we use the result of Ref.~\cite{Karagiannis:2018jdt} to model the envelope for the bispectrum theoretical error. Since our results are in redshift space, we generalize their envelope and replace the tree-level Gaussian matter bispectrum with that in redshift space. Therefore, our error envelope is anisotropic, and it is given by
\begin{equation}
E_B(k_1,k_2,k_3,\mu,\phi) = D^2(z) B_G(k_1,k_2,k_3,\mu,\phi,z) [3 b_1^3 (k_{t'}/0.15)^{1.7} + 1.8 b_2^3 k_{t'}^{1.25} + 3.2 b_1 b_2 b_3]\,,
\end{equation}
where $k_{t'} = (k_1+k_2+k_3)/3$, and for the matter expansion we use the MPTbreez scheme \cite{Bernardeau:2008fa,Crocce:2012fa} (see Appendix B of Ref.~\cite{Karagiannis:2018jdt} for a brief review). The Gaussian matter bispectrum, at tree level in perturbation theory, and in redshift space (accounting for the Kaiser term), is given by
\begin{equation}
B_G(k_1,k_2,k_3,\mu,\phi,z) = 2 Z_1({\bf k}_1,z) Z_1({\bf k}_2,z) Z_2({\bf k}_1,{\bf k}_2,z) P_0(k_1,z) P_0(k_2,z) + 2 \  {\rm perms},
\end{equation}
where $Z_i$ are the matter perturbation theory kernels in redshift-space, given by Eq.~\eqref{eq:Zfunction}, replacing $b_1 = 1$ and $b_2 = b_{K^2} = 0$. For the values of the quadratic and  third order bias, we use the fit to N-body simulations presented in Ref. \cite{Lazeyras:2015lgp}, 
\begin{eqnarray}
b_2(b_1) &=&  0.412 -  2.143 b_1 + 0.929 b_1^2 + 0.008 b_1^3, \nonumber \\
b_3(b_1) &=&  - 1.028 + 7.646 b_1 - 6.227 b_1^2 + 0.912 b_1^3,
\end{eqnarray}
and for linear bias we take $b_1 = 1.46 \sqrt{1+z}$, as we did previously.

To simplify the computation, we only consider the diagonal contribution of the error matrix, hence assuming $\rho_{ij} = \delta_{ij}$. For surveys with a wide redshift coverage, such as those considered in our work, neglecting the off-diagonal terms in the error covariance is shown to give a percent level error, and hence it is negligible for our purposes \cite{Karagiannis:2018jdt}.  We observe that the constraint on local and equilateral shapes are degraded by $10\%$ and  $20\%$, respectively. These results are in overall agreement with those in Ref.~\cite{Karagiannis:2018jdt}, in which the constraint on equilateral shape from spectroscopic surveys was shown to be degraded more than local shape. Because of differences in the modeling of the bispectrum, direct comparison is not possible. For the bispectra considered in this work, the level of degradation has a mild dependence on the spin. In general, the projected errors on $f_{\rm NL}$ are more affected compared to those on $\mu_s$. The constraints on  $f_{\rm NL}$ and $\mu_s$ are degraded by about $20\%$ and $9\%$ for $s=2$, $19\%$ and $13\%$ for $s=3$, and $17\%$ and $12\%$ for $s=4$, respectively.

\subsection{Non-Gaussian Corrections to Variance}
As described before, our main results are obtained assuming the bispectrum covariance to be Gaussian. Due to non-linearities, the covariance receives additional non-Gaussian contributions, which induce coupling between different triangular configurations (off-diagonal contributions to the covariance matrix), as well as contributions to the variance. In general it is expected that accounting for the non-Gaussian contributions to the covariance, results in weaker parameter constraints. The degradation of the constraints on a given cosmological parameter can only be fully quantified by comparing the forecasts using Gaussian variance vs. full non-Gaussian covariance, as for instance done in Ref.~\cite{Byun:2017fkz}, where non-gaussian covariances were measured from N-body simulations. 

Recently, a comparison of bispectrum covariances predicted by perturbation theory against measurement of covariances from N-body simulations was presented~\cite{Chan:2016ehg}. Neglecting redshift-space distortions, it was shown that on large scales, the dominant non-Gaussian contribution to the halo bispectrum covariance is due to fluctuations in the Poisson shot noise. Since such contributions scale as inverse powers of mean halo/galaxy number density $\bar n$, for high-density halo samples on large scales, the Gaussian covariance, if calculated using the measured (fully non-linear) halo power spectrum, is in good agreement with N-body simulations.  Based on this result, to assess the impact of NG contributions to the covariance, similar to Ref.~\cite{Karagiannis:2018jdt}, we assume that on large scales, the leading-order non-Gaussian contributions to the bispectrum variance can be well approximated by Eq.~(40) of Ref.~\cite{Karagiannis:2018jdt}. Since we perform our Fisher forecasts in redshift-space, we extend their expression to that in redshift space. We therefore consider the following expression for the covariance to account for non-Gaussian corrections to the variance in the $i^{\rm th}$ redshift bin with volume $V_i$,
\begin{align}
&\Delta B_{\rm NL}(k_1,k_2,k_3,\mu,\phi,z_i) = \frac{(2\pi)^5 s_{123}}{V_i k_1k_2k_3 \Delta k^3 \Delta \mu \Delta \phi} \left\{ \prod_{j=1}^3 \left ( P_g^{\rm obs}(k_j,\mu_j, z_i) + \frac{1}{\bar n_i} \right) \right. \\
&+ \left.  \left[ \prod_{j=1}^2 \left ( P_g^{\rm obs}(k_j,\mu_j, z_i) + \frac{1}{\bar n_i} \right) \left(P_g^{\rm obs,NL}(k_3,\mu_3,z_i) -P_g^{\rm obs}(k_3, \mu_3,z_i) + \frac{1}{\bar n_i}\right) + 2 \ {\rm perms}\right] \right\}, \nonumber
\end{align}
where $P_g^{\rm obs}$ is the observed power spectrum in redshift space, given by Eq.~(3.23), and $P_g^{\rm obs,NL}$ is obtained by replacing the linear matter power spectrum in Eq.~(3.7) by non-linear matter power spectrum, as predicted by the HALOFIT algorithm \cite{Smith:2002dz}, and calculated using the public CLASS code  \cite{Lesgourgues:2011re,Blas:2011rf}.  

Using the above expression for the bispectrum variance, we observe that our reported constraints (after imposing Planck priors), are degraded by about $35\%$ and $21\%$ for local and equilateral shapes. These results are in good agreement with the findings of Ref.~\cite{Karagiannis:2018jdt}, although direct comparison is not possible due to differences in modeling of the bispectrum. For the bispectra considered in this work, the level of degradation has a negligible dependence on the spin, and the constraints on $\mu_s$ are slightly more affected compared to the constraints on $f_{\rm NL}$. The constraints on  $f_{\rm NL}$ and $\mu$ are degraded by about $30\%$ and $33\%$ for $s=2$, $29\%$ and $33\%$ for $s=3$, and $29\%$ and $32\%$ for $s=4$.

\section{Conclusions}
\label{sec:Conclusions}
Primordial non-Gaussianity provides an invaluable window into the physics of the very-early Universe, at energy scales beyond the reach of any particle collider. Even in the simplest scenario of inflation driven by a single scalar field, there could be additional particles present during inflation. The presence of these particles leaves an imprint on late-time correlation functions of curvature fluctuations. In particular, particles with masses of the order of the Hubble scale during inflation produce a distinct shape in the primordial bispectrum with an angular dependence determined by the spin of the particles. Furthermore, one of the contributions to the produced bispectrum has an oscillatory behavior, the frequency of which is determined by the mass of the particles. 

We have computed a template that captures the non-Gaussian signature of these new particles, which can be used for particles with spins $s=2, 3, 4$ over a wide range of masses parametrized by the parameter $\nu$. 
This template is valid for a general triangle configuration and we have tested it against the full numerical calculation of the bispectrum. Using this template, we made a forecast for how well these signatures can be measured with next-generation galaxy surveys, in particular we considered DESI and EUCLID. We modeled the galaxy bispectrum in redshift space using tree-level perturbation theory and we accounted for suppression of the bispectrum due to the Finger-of-God and Alcock-Paczynski effects. We obtained constraints on parameters of primordial non-Gaussianity, marginalizing over the cosmological parameters and biases.

For particles with masses comparable to the Hubble scale during inflation, our results indicate that any non-Gaussianity larger than $f_{\rm NL}  \gtrsim 1 $ should be observable with DESI or EUCLID. We have found that, if such non-Gaussianity was detected, the mass of the spinning particle (or more precisely, the parameter $\nu$ defined in Eq.~\eqref{eq:nu}) could be measured to at least tens-of-percent precision. This shows the degree to which these new shapes of non-Gaussianity are unique, and it provides us with a smoking gun of the presence of massive particles during inflation. This work is the first forecast on constraints on massive spinning particles through the galaxy bispectrum, complementing previous constraints from the galaxy power spectrum~\cite{MoradinezhadDizgah:2017szk}, constraints for extra scalar degrees of freedom from power spectrum~\cite{Sefusatti:2012ye,Norena:2012yi,Gleyzes:2016tdh}, and the bispectra of other observables~\cite{Meerburg:2016zdz}.

We conclude by noting that, given the forecasted detectability of the signatures of spinning particles from the galaxy bispectrum, it is important to improve upon the model of galaxy bispectrum to go beyond the leading-order in perturbation theory. This would require accounting for loop contributions from gravitational evolution of matter as well as those due to primordial non-Gaussianity. 
Moreover, the Gaussian assumption for the bispectrum covariance needs to be tested and improved. We provide an estimate for the degradation of the forecasted constraints due to the uncertainty in theoretical modeling of the bispectrum and the leading-order non-Gaussian corrections to the bispectrum variance. We found that, on average, for particles with spins $s=2,3,4$, accounting for the theoretical errors degrade the constraints on $f_{\rm NL}$ by about $20\%$, while the constraints on $\mu$ are degraded by about $10\%$. The constraints seem to be slightly more sensitive to the non-Gaussian corrections to the bispectrum variance, as accounting for the leading contribution degrades the constraints on $f_{\rm NL}$ and $\mu$  by about $30\%$ and $33\%$, respectively. 

\paragraph{Acknowledgements.} 
It is our pleasure to thank Emiliano Sefusatti, Vincent Desjacques and Roman Scoccimarro for helpful discussions.
This work was supported by the Dean's Competitive Fund for Promising Scholarship at Harvard University.

\appendix
\section{Bispectrum Template}\label{app:template}
In this appendix, we derive the bispectrum template that was presented in the main text. In Section \ref{app:inin}, we give details of the in-in computations of the bispectrum involving the exchange of a massive spin-$s$ field. 
We fix the relative amplitude between the analytic and non-analytic parts in Section \ref{app:procedure}.

\subsection{In-In Results}\label{app:inin}
The bispectrum involving a single massive spinning particle exchange is
\begin{align}
\langle\zeta_{\k_1}\zeta_{\k_2}\zeta_{\k_3}\rangle' \propto P_s(\hat \k_1\cdot\hat\k_3)\, {\cal I}^{(s)}(k_1,k_2,k_3) + \text{5 perms}\, ,\label{zeta3spinssingle}
\end{align}
where we have only kept the momentum and mass dependences. The integral representation of the function ${\cal I}^{(s)}$ is given by
\begin{align}
{\cal I}^{(s)}&\equiv \frac{2\pi^3 {\cal N}_s^2}{k_1^{3/2} k_2^{7/2} k_3}{\rm Re}\big[\I_{1}^{(s)}+\I_{2}^{(s)}+\I_{3}^{(s)}\big]\, ,\\[4pt]
	\I_1^{(s)}&\equiv \int_0^\infty dx\, x^{s-1/2}(1-ix)\G^{(s)}_{i\nu}(\kappa_{12} x)e^{ix(1+\kappa_{32})}\int_0^\infty d y\, y^{s-5/2}(1+iy)\G^{(s)*}_{i\nu}(y)e^{-iy}\, ,\label{I1} \\[2pt]
	\I_2^{(s)} &= \int_0^\infty d x\, x^{s-1/2}(1+ix)\G^{(s)}_{i\nu}(\kappa_{12} x)e^{-ix(1+\kappa_{32})}\int_{\kappa_{12}x}^\infty d y\, y^{s-5/2}(1+iy)\G^{(s)*}_{i\nu}(y)e^{-iy}\, ,\label{I2}\\[2pt]
	\I_3^{(s)} &= \int_0^\infty d x\, x^{s-5/2}(1+ix)\G^{(s)}_{i\nu}(x)e^{-ix}\int_{\kappa_{21}x}^\infty d y\, y^{s-1/2}(1+iy)\G_{i\nu}^{(s)*}(\kappa_{12} y)e^{-iy(1+\kappa_{32})}\, ,\label{I3}
	\end{align}
where $\kappa_{ij}\equiv k_i/k_j$, 
\begin{align}
{\cal N}_s^2 &\equiv  \frac{\pi^2}{4}\frac{s!}{(2s-1)!!}\frac{{\rm sech}\,\pi\nu}{\Gamma(\frac{1}{2}+s+i\nu)\Gamma(\frac{1}{2}+s-i\nu)}\, , \label{Ns}
\end{align}
is the normalization constant, and $\G^{(s)}_{i\nu}$ is the mode function of a spin-$s$ field with the overall time scaling factored out. Ignoring an irrelevant phase, we have~\cite{Lee:2016vti}
\begin{align}
	\G_{i\nu}^{(2)}(x)& \equiv \frac{e^{-\pi\nu/2}}{12}\hskip 1pt \Big[{\mp}6x(2\pm i\nu)H_{i\nu\pm 1}(x)+(8x^2-9)H_{i\nu}(x)\Big]\, ,\\
	\G_{i\nu}^{(3)}(x)& \equiv \frac{e^{-\pi\nu/2}}{120}\Big[{\mp}x(4x^2+345+30\nu(2\nu\pm 9i))H_{i\nu\pm 1}(x)+ (244x^2-225)H_{i\nu}(x)\Big],\\
	\G_{i\nu}^{(4)}(x)& \equiv \frac{e^{-\pi\nu/2}}{1680}\Big[{\mp} 4x\big(2x^2(44\mp 83i\nu)-105i(\mp 4i+\nu)(-11+2\nu(\mp 4i+\nu)\big)H_{i\nu\pm 1}(x)\nonumber\\
	&\qquad \qquad\ \ \, {+}(704x^4+24x^2(571-70\nu^2)-11025)H_{i\nu}(x)\Big]\, ,
\end{align}
where we have left the sum over $\pm$ implicit and casted the expressions in terms of $H_{i\nu\pm 1}$ and $H_{i\nu}$ by the use of the recurrence relation, $H_{n+1}(x)+H_{n-1}(x)=\frac{2n+1}{x}H_{n}(x)$.

\paragraph{Non-analytic part.}
For the non-analytic part, we will quote the result from~\cite{Lee:2016vti}. The non-analytic part of the bispectrum in the squeezed limit is
\begin{align}
	\lim_{k_1\ll k_3}\langle\zeta_{\k_1}\zeta_{\k_2}\zeta_{\k_3}\rangle'&=\frac{A^{(s)}}{k_1^3k_3^3}\left(\frac{k_1}{k_3}\right)^{3/2}\cos\bigg[\nu\ln\frac{k_1}{k_3}+\phi_s\bigg]P_s(\hat\k_1\cdot\hat\k_3) +(\k_2\leftrightarrow\k_3)\, ,
\end{align}
where
\begin{align}
	A^{(s)} \equiv \frac{\pi^{7/2}s\hs f^{(s)}(\nu)}{2^{2s+1}(2s-1)!!}\frac{\Gamma(\frac{1}{2}+s-i\nu)\Gamma(\frac{1}{2}+s+i\nu)}{(s-\frac{3}{2})^2+\nu^2}\bigg[\Big((5+2s)^2+4\nu^2\Big)\,\frac{{\rm coth}\,\pi\nu}{\nu}\bigg]^{1/2}\, .\label{Aosc}
\end{align}
The mode function of spinning fields can be obtained using a recursive formula, and the function $f^{(s)}(\nu)$ represents the ratio of the integrals involving the quadratic vertex evaluated using the exact mode function and a simple Hankel function, defined by
\begin{align}
	f^{(s)}(\nu) = \frac{ \int_0^\infty d x\, x^{s-5/2}(1-ix)\G_{i\nu}^{(s)}(x)e^{ix}}{e^{-\pi\nu/2}\int_0^\infty d x\, x^{s-5/2}(1-ix)\H_{i\nu}(x)e^{ix}}\, .\label{fs}
\end{align}
This can be computed for each spin, e.g.
\begin{align}
	f^{(2)} &= -\frac{985-664\nu^2+16\nu^4}{576}\, ,\\
	f^{(3)} &= \frac{4800519-2564556\nu^2+96366\nu^4-64\nu^6}{460800}\, , \\
	f^{(4)} &= \frac{-221842845+341268176\nu^2-39893856\nu^4+278784\nu^6+2816\nu^8}{12902400}\, .
\end{align}
In general, $f^{(s)}$ is a degree-$2s$ polynomial in $\nu$.

\paragraph{Analytic part.}
For extracting the analytic part of the full bispectrum, we compute the bispectrum due to a local operator in Eq.~\eqref{Lpi}. 
The contribution of the coupling \eqref{Lpi} to the bispectrum is given by
\begin{align}
	\langle \zeta_{\k_1}\zeta_{\k_2}\zeta_{\k_3}\rangle'
	& \propto \frac{P_s(\hat \k_2\cdot\hat \k_3)}{k_1(k_2k_3)^{3-s}k_t^{2s+1}}\Big[(2s-1)\big((k_2+k_3)k_t+2sk_2k_3\big)+k_t^2 \Big] + \text{2 perms}\, .\label{Qbispectrum}
\end{align}
We match the amplitude of this shape to that of the analytic part of the full bispectrum involving the massive particle exchange (through the procedure which we outline in the next section).

\subsection{Derivation}\label{app:procedure}

Unlike the non-analytic part, the integrals corresponding to the analytic part of the bispectrum do not factorize. While this makes it difficult to perform the integration for arbitrary momenta by analytic means, a closed-form expression of the resulting bispectrum can still be obtained in the squeezed limit in a straightforward way as follows.

\paragraph{Preliminaries.} 
In order to obtain the final expressions, we will need a few ingredients. Let us consider a simpler version of the integral~\eqref{I2} given by

\begin{align}
	\int_0^\infty d x\, x^{s-1/2+|\tilde\epsilon|}\H_{i\nu+\tilde\epsilon}(\kappa_{12} x)e^{-ix(1+\kappa_{32})}\int_{\kappa_{12}x}^\infty d y\, y^{t-5/2+|\epsilon|}\H^*_{i\nu+\epsilon}(y)e^{-iy}\, ,\label{I2simple}
\end{align}
where the mode function $\G_{i\nu}^{(s)}$ is replaced by the Hankel function of the first kind $x^{|\epsilon|}\H_{i\nu+\epsilon}$ with $\tilde\epsilon,\epsilon\in\{-1,0,1\}$. The indefinite integral of the inner layer can be done analytically, which is given by
\begin{align}
&\J_{i\nu+\epsilon}(t,x) \equiv \int  d x\, x^{t-\frac{5}{2}+|\epsilon|}\hs \H_{i\nu+\epsilon}^*(x)e^{-ix}\approx\frac{ie^{\pi\nu}\hs 2^{i\nu}x^{t-\frac{3}{2}-i\nu}\Gamma(i\nu)}{\pi(t-\frac{3}{2}-i\nu)}\nonumber\\
&\times \left\{ {}_2\bhs F_2\bigg[\begin{array}{c}\frac{1}{2}-i\nu,t-\frac{3}{2}-i\nu\\ t-\frac{1}{2}-i\nu,1-2i\nu\end{array};-2ix\bigg]\delta_0^\epsilon-{}_2\bhs F_2\bigg[\begin{array}{c}-\frac{1}{2}-i\nu,t-\frac{3}{2}-i\nu\\ t-\frac{1}{2}-i\nu,1-2i\nu\end{array};-2ix\bigg]2i\nu\hs\delta_{{-\! 1} }^\epsilon\right\}+\cdots \, ,
\end{align}
where the approximation sign and ellipses indicate that terms that are suppressed by factors of $e^{-\pi\nu}$ and with subleading powers in $x$ are being dropped, respectively, and
\begin{align}
	{}_m\bhs F_n\bigg[\begin{array}{c}a_1,\cdots, a_m\\ b_1,\cdots, b_n\end{array};x\bigg]\equiv \sum_{k=0}^\infty\frac{(a_1)_k\cdots(a_m)_k}{(b_1)_k\cdots(b_n)_k}\frac{x^k}{k!}\, ,
\end{align}
is the generalized hypergeometric function with $(a)_k\equiv \Gamma(a+k)/\Gamma(k)$ being the Pochhammer symbol. In the $\kappa_{12}\to 0$ limit, we can expand the Hankel function in the outer layer
\begin{align}
	x^{|\epsilon|}\H_{i\nu+\epsilon}(x\to 0)\, & \approx\, -i\big(\delta^\epsilon_0+2i\nu\hs \delta^\epsilon_{-1} \big)\frac{e^{\pi\nu}\Gamma(-i\nu)}{\pi}\left(\frac{x}{2}\right)^{i\nu} \equiv \, {\sf h}_\epsilon\hs x^{i\nu}\, .
\end{align}
Note that only the $x^{i\nu}$ term becomes relevant for large $\nu$. We then consider the outer integral
\begin{align}
	&\K_{i\nu+\epsilon}(s,t,p,q) \equiv p^{i\nu}\int_0^\infty  d x\, x^{s-\frac{1}{2} +i\nu}\J_{i\nu+\epsilon}(t,p x)e^{-iqx}\nonumber\\
	&\approx \frac{ e^{\pi\nu}}{\sinh\pi\nu}\frac{\left(p/q\right)^{t-1}}{p^{1/2}q^s}\frac{2^{i\nu}i^{-1-s-t}}{t-\frac{3}{2}-i\nu}\frac{\Gamma(s+t-1)}{\Gamma(1+i\nu)} \times \Bigg\{ {}_3\bhs F_2\Bigg[\begin{array}{c} \frac{1}{2}-i\nu,\hs t-\frac{3}{2}-i\nu,\hs s+t-1\\t-\frac{1}{2}-i\nu,1-2i\nu\end{array}\hs;-\frac{2p}{q}\Bigg]\delta^\epsilon_0\nonumber\\
	& - {}_3\bhs F_2\Bigg[\begin{array}{c} -\frac{1}{2}-i\nu,\hs t-\frac{3}{2}-i\nu,\hs s+t-1\\t-\frac{1}{2}-i\nu,1-2i\nu\end{array}\hs;-\frac{2p}{q}\Bigg]2i\nu\delta^\epsilon_{-\! 1}\Bigg\}+\cdots \, .\label{Kint2}
\end{align}
To arrive at this result, we have performed the integration of the series element of the hypergeometric function and then taken the resummation. Note that the hypergeometric function in Eq.~\eqref{Kint2} simply becomes unity in the small $p$ limit. 

\paragraph{Results.} 
Using the above formulas, we will now derive the amplitudes of the analytic part of the bispectrum for different spins. First, consider the integral with the proper spin-2 mode function
\begin{align}
	{\cal I}_2^{(2)} = \int_0^\infty d x\, x^{s-1/2}(1+ix)\G^{(2)}_{i\nu}(\kappa_{12} x)e^{-ix(1+\kappa_{32})}\int_{\kappa_{12}x}^\infty d y\, y^{s-5/2}(1+iy)\G^{(2)*}_{i\nu}(y)e^{-iy}\,.
\end{align}
In the late-time limit, the spin-2 mode function becomes
\begin{align}
	\G_{i\nu}^{(2)}(x\to 0) &= \frac{e^{-\pi\nu/2}}{12}\big(6(2-i\nu){\sf h}_{-1}-9{\sf h}_0\big) x^{i\nu} \equiv {\sf g}^{(2)} x^{i\nu}\,.
\end{align}
The outer integrand simplifies, and the integral takes the form
\begin{align}
	\K^{(2)} (s,t,p,q) &\equiv p^{i\nu}\int_0^\infty  d x\, x^{s-\frac{1}{2}+i\nu}e^{-iqx}\int^\infty_{p x}  d y\, y^{t-\frac{5}{2}}\hs \G^{(2)*}_{i\nu}(y)e^{-iy}\nonumber\\
	&\approx \frac{e^{-\pi\nu/2}}{12}\Big[6(2+i\nu)\K_{i\nu-1}(s,t,p,q)-9\K_{i\nu}(s,t,p,q)\Big]\, .
\end{align}
The full integral \eqref{I2} is then given by
\begin{align}
	{\cal I}^{(2)}_2 = {\sf g}^{(2)}\sum_{\epsilon,\tilde\epsilon=0,1} i^{\epsilon+\tilde\epsilon}\hs \K^{(2)}(s+\epsilon,s+\tilde\epsilon,\kappa_{12},2) \, .\label{ReI2}
\end{align}
In the squeezed limit, this becomes
\begin{align}
	{\rm Re}[{\cal I}_2^{(2)}] \approx \sqrt{\kappa_{12}}\,\frac{5(9+4\nu^2)}{16\pi}\, .
\end{align}
Following a similar analysis for higher spins, we find
\begin{align}
	{\rm Re}[{\cal I}_2^{(3)}] &\approx -21\kappa_{12}^{3/2}\,\frac{225+1036\nu^2+2032\nu^4+64\nu^6}{32\pi(9+4\nu^2)}\, , \\
	{\rm Re}[{\cal I}_2^{(4)}] & \approx 405\kappa_{12}^{5/2}\,\frac{11025+196144\nu^2+24864\nu^4+5376\nu^6+256\nu^8}{512\pi(25+4\nu^2)}\, .\label{ReI4}
\end{align}
Finally, the relative amplitudes between the analytic and non-analytic parts for different spins defined in Eq. \eqref{BNA} are given by comparing Eqs.\eqref{ReI2}-\eqref{ReI4} with Eq.~\eqref{Aosc}. The results are
\begin{align}
	r^{(2)} &= \frac{\pi^{9/2}(1+4\nu^2)(9+4\nu^2)f^{(2)}(\nu)}{256(1+\coth\pi\nu)}\sqrt{\frac{2(81+4\nu^2)}{\nu\sinh 2\pi\nu}}\, , \\[2pt] 
	r^{(3)} &=-\frac{5\pi^{9/2}[(1+4\nu^2)(9+4\nu^2)(25+4\nu^2)]^2f^{(3)}(\nu)}{688128(225+1036\nu^2+2032\nu^4+64\nu^6)}\sqrt{\frac{2(121+4\nu^2)}{\nu\sinh 2\pi\nu}}\, ,\\[2pt]
	r^{(4)} &=\frac{\pi^{9/2}[(1+4\nu^2)(9+4\nu^2)(25+4\nu^2)(49+4\nu^2)]^2f^{(4)}(\nu)}{31850496(11025+196144\nu^2+24864\nu^4+5376\nu^6+256\nu^8)}\sqrt{\frac{2(169+4\nu^2)}{\nu\sinh 2\pi\nu}}\, .
\end{align}
These amplitudes scale as $e^{-\pi\nu}$ in the large-$\nu$ limit, as expected. In deriving these formulas, we have dropped terms that are further suppressed by factors of $e^{-\pi\nu}$. 

\section{Shape Functions: Extended}
\label{sec:AppShapes}

In this appendix, we elaborate on the different shape functions we have chosen and their characteristics.

\subsection{Angular Dependence}
\begin{figure}
	\includegraphics[height=0.235\linewidth]{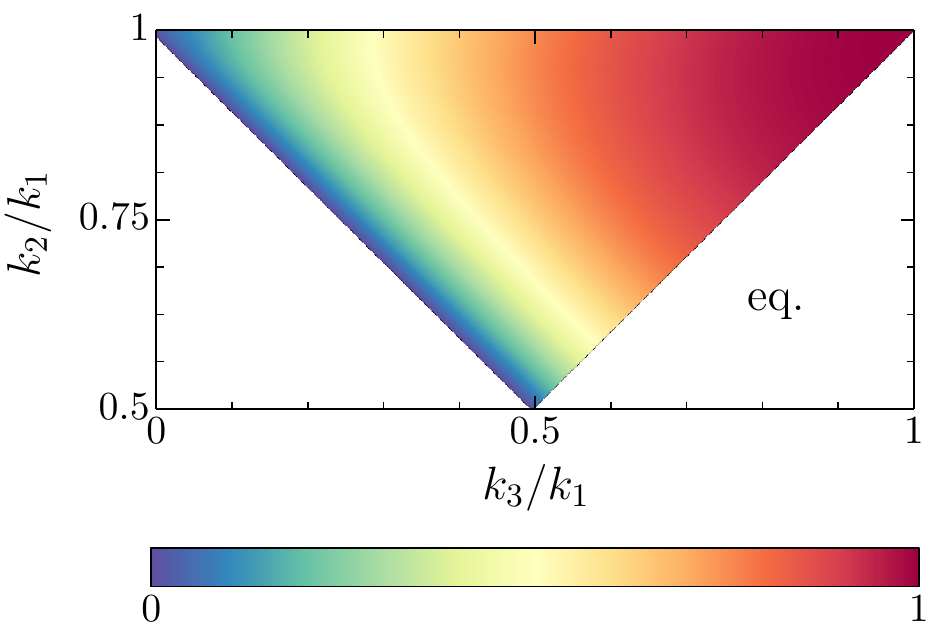}
	\includegraphics[height=0.235\linewidth]{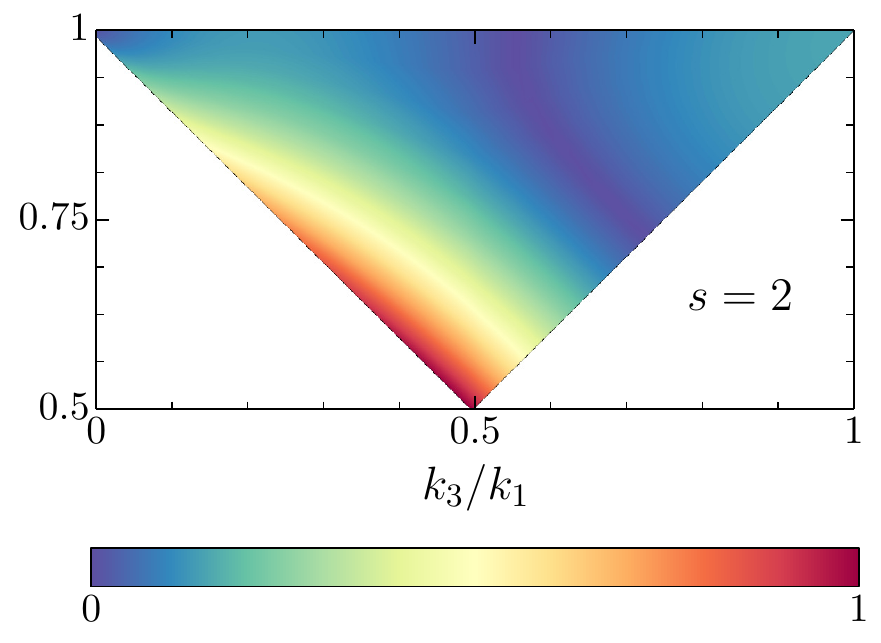}	
	\includegraphics[height=0.235\linewidth]{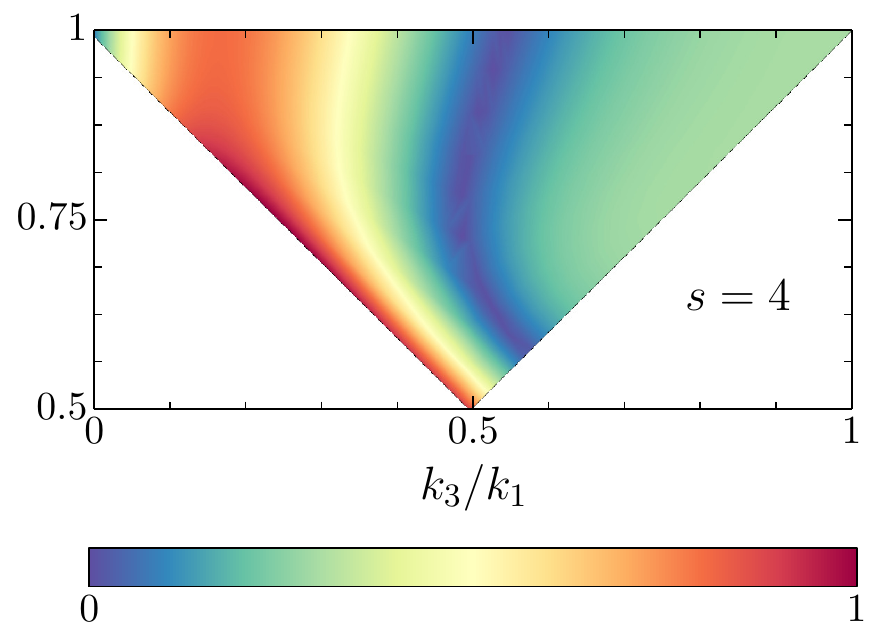}	
	\caption{Shape functions, normalized to unity at their maximum, as a function of the $k_2/k_1$ and $k_3/k_1$ ratios, while keeping $k_1$ fixed. From left to right we show equilateral-type non-Gaussianities, and our template for $s=2$ and $s=4$, both in the $\nu\gg1$ limit.}
	\label{fig:shapes}
\end{figure}

In Section~\ref{sec:Primordial}, we defined the shape of primordial non-Gaussianities, both for the usual shapes and for our template.
Given that the overall scaling is the same for all shapes, we will fix $k_1$ and vary $k_2, k_3<k_1$, choosing the three $k$-values to form a triangle. We show the shape functions, using this parametrization, for equilateral-type non-Gaussianities, as well as the analytic ($\nu\gg1$) limit of our template for $s=2$ and 4 in Fig.~\ref{fig:shapes}.
From these panels, it is visually clear that these three cases can all be distinguished from each other, given their different angular structures.

\begin{figure}	
	\includegraphics[height=0.24\linewidth]{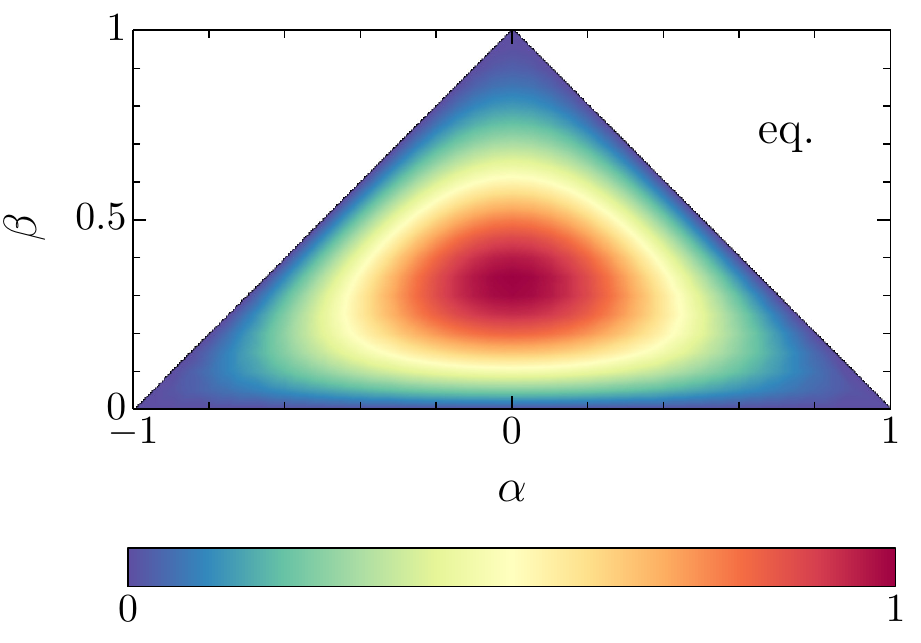}
	\includegraphics[height=0.24\linewidth]{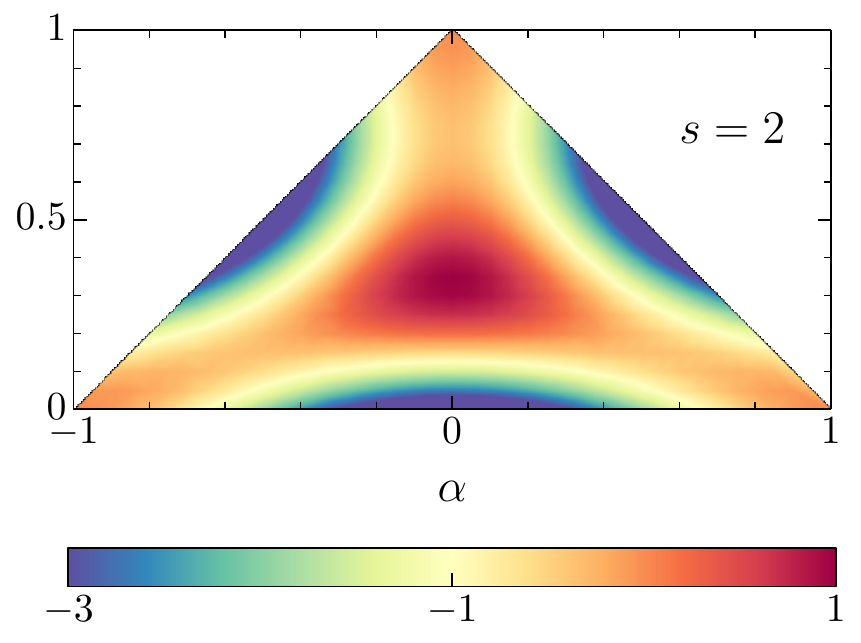}	
	\includegraphics[height=0.24\linewidth]{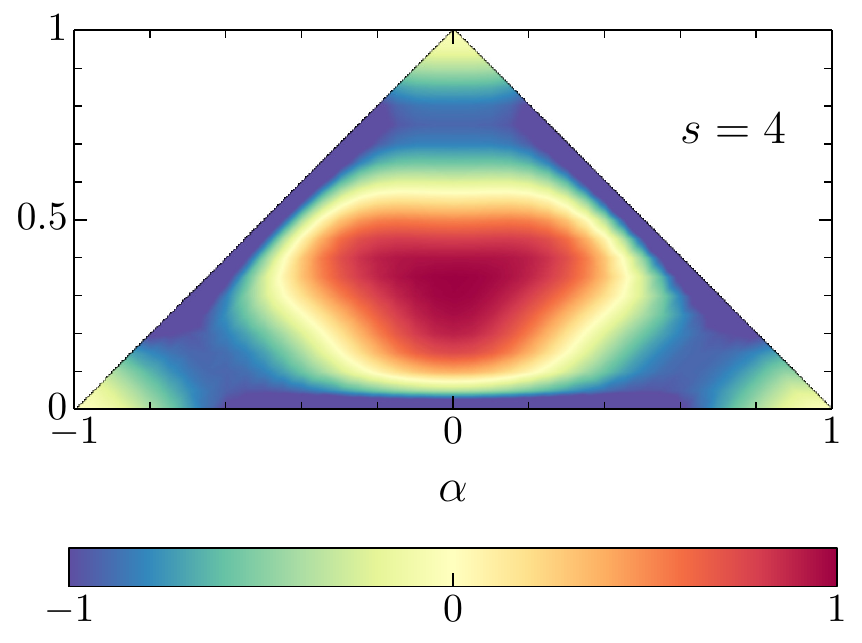}	
	\caption{ Shape functions, with the variables defined in Eq. \eqref{eq:alpha_beta}, divided by the local one, for equilateral non-Gaussanities, as well as for the spin $s=2$ and $s=4$ cases. By construction, all shapes have the same value in the equilateral case.
	}
	\label{fig:shape_ab}
\end{figure}
To further illustrate the difference between the types of non-Gaussianity, we will now describe the triangles in terms of two new variables, $\alpha$ and $\beta$, following Ref.~\cite{Fergusson:2008ra}, as
\begin{align}
k_1 &= k_t(1-\beta)\, , \quad k_2 = k_t(1+\alpha+\beta)/2\, , \quad k_3 = k_t(1-\alpha+\beta)/2\, ,
\label{eq:alpha_beta}
\end{align}
where the overall scale $k_t$ is irrelevant.  The variables $\alpha$ and $\beta$ run between $0\leq\beta\leq1$, and $-(1-\beta)\leq\alpha\leq 1-\beta$. Heuristically, these variables can be understood to parametrize the triangle resulting from causing a section of the polyhedron formed by the triangle identities, at fixed $k_t$. 
We show in Fig.~\ref{fig:shape_ab} the ratio of different shapes divided by the local shape of non-Gaussianities, where we have again set $\nu\gg1$ for the spinning cases. The local shape is expected to be larger for squeezed configurations, at the edge of the triangles in Fig.~\ref{fig:shape_ab}. This is clear for the equilateral case, and is less pronounced for the spinning cases.
Moreover, the spinning cases show a richer angular dependence, which will allow us to distinguish different spins.

\subsection[High-$\nu$ Correlator]{High-$\boldsymbol{\nu}$ Correlator}

In Section \ref{sec:correlations}, we explored the shape correlation between the shapes of primordial non-Gaussianities arising from $s=2$, 3 and 4 particles, focusing in the $\nu=3$ and $\nu=6$ cases. We will now perform the same calculation for other spins, limiting ourselves to the $\nu\gg1$ limit, where only the analytic part of the template in Eq.~\eqref{eq:Bfull} survives.

\begin{figure}
\centering	
	\includegraphics[scale=1.3]{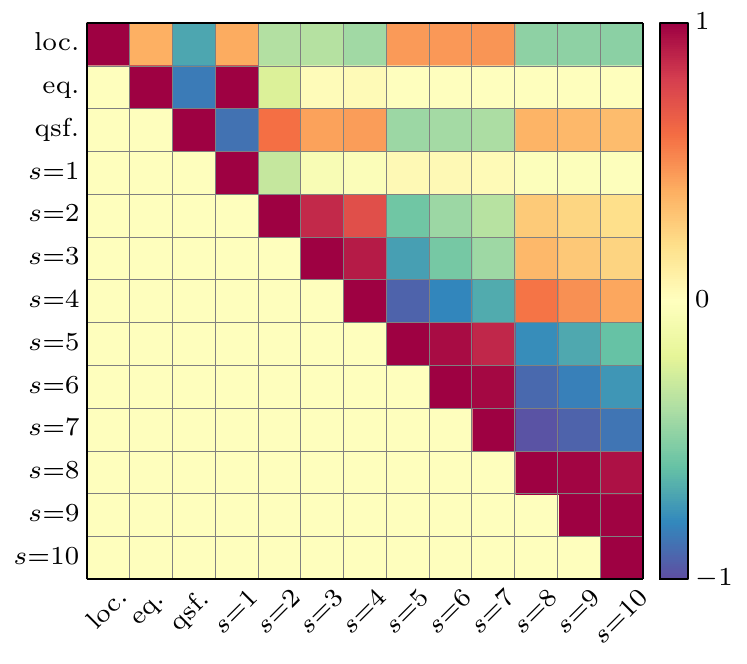}	
	\caption{Correlation between shapes, as defined in Eq. \eqref{eq:CorrPrim}, for the high-$\nu$ case of spinning particle, as well as the local, equilateral and quasi-single-field shapes, for comparison. }
	\label{fig:corr_highmu}
\end{figure}
We compute the correlation between shapes as in Eq.~\eqref{eq:CorrPrim},
and we show it in Fig.~\ref{fig:corr_highmu}, where we also compare with the usual local, equilateral, and quasi-single field shapes.
Interestingly, we see that higher spins are more correlated with their neighbors. By finding the spin change $\Delta s$ where the correlation length drops by a factor of 2, we find a ``correlation length" in spin space of $\Delta s \approx 2 s$, increasing at higher spins. This result is complementary to Fig.~\ref{fig:corr}, where the ``oscillatory" part of the template in Eq. \eqref{eq:Bfull} breaks the degeneracy between neighboring spins.

\section{Power-Spectrum Constraints}
\label{sec:PowSp}

In addition to a contribution to the galaxy bispectrum, PNG also leaves an imprint on the power spectrum by inducing a scale-dependent contribution to the bias \cite{Dalal:2007cu,Matarrese:2008nc,Afshordi:2008ru,Slosar:2008hx}.  In this section, we explore the constraints that can be achieved with the galaxy power spectrum, which complements our forecast from bispectrum. We consider only the EUCLID survey for this analysis.

In our forecast, we use the result of \cite{Desjacques:2011jb,Desjacques:2011mq} (see also \cite{Scoccimarro:2011pz}), in which a general expression for the scale-dependent correction to linear bias due to primordial non-Gaussianity is derived. Accounting only for the contribution from primordial bispectrum, their expression reduces to
\be\label{eq:deltab_NG}
\Delta b_1^{\rm NG}(k,z) =  \frac{2 \mathcal F_R^{(3)}(k,z)}{\mathcal M_R(k,z)} \left[(b_1-1)\delta_c + \frac{d \ln {\mathcal F}^{(3)}_R(k,z)}{d\ln \sigma_R}\right],
 \ee
where $\delta_c=1.686$, is the threshold of spherical collapse, ${\mathcal F}^{(3)}_R(k,z)$ is the shape factor defined as
\be \label{eq:F3_R}
{\mathcal F}^{(3)}_R(k,z) = \frac{1}{4 \sigma^2_{R}(z) P_\zeta(k)} \int \frac{d^3 q}{(2\pi)^3} \mathcal M_R(q,z) \mathcal M_R(|\bk-\bq|,z) B_\zeta(-\bk,\bq, \bk-\bq),
\ee
and $\sigma_R$ is the variance of the density field smoothed over a scale $R(M) = (3M/4 \pi \bar \rho)^{1/3}$,
\be
\sigma_R^2(z) = \int_0^\infty \frac{dk }{2\pi^2} k^2 P_\zeta(k) \mathcal M_R^2(k,z),
\ee
with $\mathcal M_R(k,z) = W_R(k)\mathcal M(k,z)$, where $W_R(k)$ is the Fourier transform of a spherical tophat filter with radius R,
\be
W_R(k) = \frac{3\left[{\rm sin}(kR) - kR \  {\rm cos} (kR)\right]}{(kR)^3}. 
\ee
We refer the interested reader to the references above for details of the derivation of the above result and its physical interpretation, and only discuss our forecasts here.

\begin{table*}[t!]
	\centering
	\begin{tabular}{| l || c | c || c | c |}
		\hline
		EUCLID &  \multicolumn{2}{c||}{Bispectrum} & \multicolumn{2}{c|}{Power Spectrum}\\
		\hline \hline
		& $\sigma(f_{\rm NL})$ & $\sigma(\tilde \nu)$ & $\sigma(f_{\rm NL})$ & $\sigma(\tilde \nu)$
		\\
		\hline
		loc.	& 0.38	& $-$	& 3.3	& $-$
\\
eq.	& 2.3	&$-$	&  70	& $-$
\\
qsf ($\tilde \nu_{\rm fid}=1$)	&  2.2	& 1.3	& 38	& 21
\\
\hline \hline
$\nu_{\rm fid}=3$ & $\sigma(f_{\rm NL})$ & $\sigma(\nu)$ & $\sigma(f_{\rm NL})$ & $\sigma(\nu)$	
\\ \hline		
$s=2$	& 0.66	& 0.35	& 23	& 84
\\
$s=3$	& 1.5	& 3.3	& 71	& 230
\\
$s=4$	& 0.68	& 0.098	& 28	& 7.3
\\	
\hline
\end{tabular}
	\caption{1-$\sigma$ uncertainties in non-Gaussianity parameters for different models from bispectrum and power spectrum measured by EUCLID survey. The constraints are obtained for $f_{\rm NL}^{\rm fid} = 1$, $\nu_{\rm fid}=3$, $\tilde \nu_{\rm fid}=1$, assuming $k_{\rm max}(z=0)=0.15 \ h \, {\rm Mpc}^{-1}$, and marginalizing over all other parameters.}		
	\label{tab:fNLPowSp}
\end{table*}

As expected, the constraints from the power spectrum are always weaker than those derived from the bispectrum. For the local shape, since the bias receives a distinct $1/k^2$-dependence, order unity constraints on $f_{\rm NL}$ is achievable. Nonetheless, the bispectrum improves upon these constraints since it probes more modes. For equilateral and QSF models as well as higher-spinning case, the constraints from bispectrum are significantly stronger than those from the power spectrum.  Among various spins, the constraint on spin-3 particles is the weakest.

\begin{figure}[t]
\hskip 30pt
	\includegraphics[width=0.8 \textwidth]{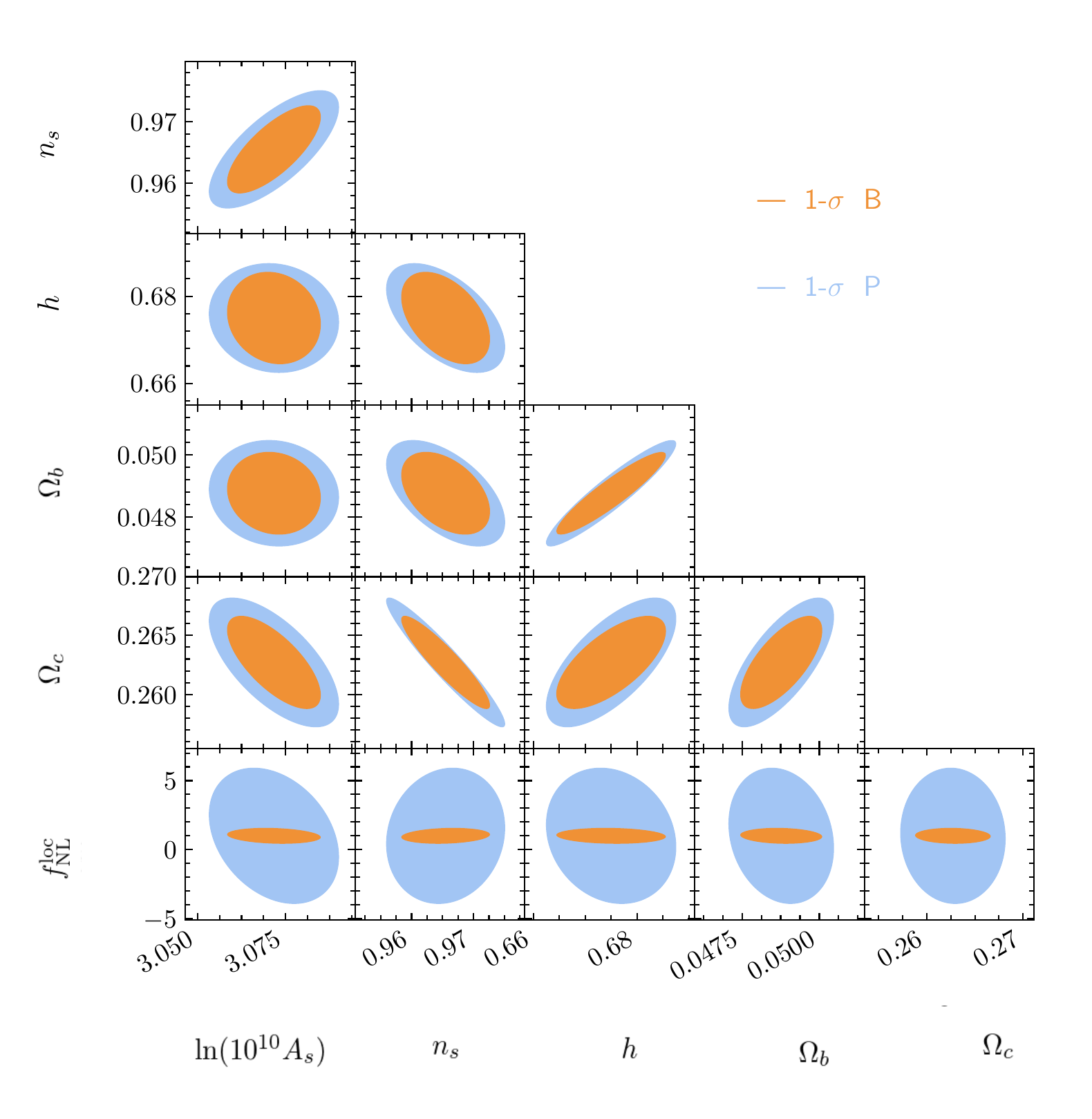}	
	\caption{1-$\sigma$ confidence ellipses in the five $\Lambda$CDM parameters, as well as the amplitude of primordial non-Gaussianity of local shape, for EUCLID.
	Our fiducial values are the same as in Table~\ref{tab:fNLE}, which for the non-Gaussianity parameters correspond to $f_{\rm NL}=1$.	
	The orange (inner) contours correspond to the bispectrum-only constraints, while the blue (outer) ones are obtained from power spectrum only. The contours are obtained using $k_{\rm max} = 0.15 \ h \, {\rm Mpc}^{-1}$.}
	\label{fig:Ellipses_PS}
\end{figure}

To visually compare the power spectrum vs.~bispectrum constraints, we show in Fig.~\ref{fig:Ellipses_PS} the 1-$\sigma$ ellipses for cosmological parameters and $f_{\rm NL}$ for the local shape PNG. This plot is made choosing $k_{\rm max} = 0.15 \ h\, {\rm Mpc}^{-1}$ at $z=0$, and using Eq.~\eqref{eq:kmax} for other redshifts.  

\bibliographystyle{utphys}
\bibliography{Bispectrum}

\end{document}